\documentclass{ocephys}
\let\linenumbers\relax

\newcommand{\xk}{\vec{x}_\mathrm{K}}
\newcommand{\xn}{\vec{x}_\mathrm{N}}

\newcommand{\T}{^\mathrm{T}}

\DeclareSIUnit{\cpkm}{cpkm}

\usepackage{xr}
\usepackage{lineno}

\title{Isolating Balanced Ocean Dynamics \\ in SWOT Data}

\author[1,2]{J.~W.~Skinner}
\author[2]{J.~Callies}
\author[1]{A.~Lawrence}
\author[2]{X.~Zhang}

\affil[1]{Brandeis University, Waltham, MA}
\affil[2]{California Institute of Technology, Pasadena, CA}

\corr{J.~W.~Skinner, \href{mailto:jwskinner@ucsd.edu}{jwskinner@ucsd.edu}}

\begin{document}

\section*{Key Points}

\begin{itemize}
  \item We isolate balanced sea surface height (SSH) signals in SWOT measurements using a Bayesian inversion based on Gaussian processes.
  \item The method fills the nadir gap and extracts meso- and submesoscale flow structures from originally noisy SSH maps.
  \item Tests on synthetic data generated from a high-resolution simulation illustrate the reliability of the extraction and its estimated uncertainty.
\end{itemize}

\section*{Abstract}

The Surface Water and Ocean Topography (SWOT) mission provides two-dimensional sea surface height anomaly (SSHA) maps at unprecedented resolution, but its signal is a combination of balanced meso- and submesoscale turbulence, unbalanced motions such as internal waves, and small-scale noise.
Interpreting the meso- and submesoscale flow features captured by SWOT requires careful isolation of the balanced signal. We present a statistical method to do so in regions where internal tides make a negligible contribution to the SSHA signal, such as western boundary current regions and the Southern Ocean.
Our method assumes Gaussian statistics as a Bayesian prior for both the balanced flow and the remainder of the signal, and we infer their respective covariance functions by fitting parametric models to the observed SSHA wavenumber spectrum. Using these inferred covariance functions, we perform a Bayesian inversion to produce swath-aligned SSHA maps that isolate the balanced signal and fill the nadir gap.
We evaluate the method using synthetic data from a high-resolution simulation with realistic SWOT-like small-scale noise added. Comparisons with the underlying model data show that our inversion successfully removes the unbalanced small-scale signal while preserving meso- and submesoscale eddies, fronts, and filaments down to a spatial feature scale (the inverse angular wavenumber) of about \qty{6}{\kilo\meter} in the subtropical Northwest Atlantic region used for illustration. This comparison also demonstrates that the posterior uncertainty is a reliable estimate of the error. While our focus is on balanced dynamics, the balanced extraction can be extended to include tides, internal waves, and non-Gaussian priors.

\section*{Plain Language Summary}

Ocean dynamics in the \qtyrange{1}{100}{\kilo\meter} range are central to the transport of heat and carbon in the upper ocean. 
The recently launched SWOT mission measures sea surface height at high resolution and in two dimensions, revealing eddies, fronts, and filaments that characterize turbulent ocean dynamics at these scales in greater detail than previously possible. 
However, SWOT measurements also contain small-scale signals not clearly linked to this turbulent circulation, which complicates the interpretation of the data.
Treating these small-scale signals as noise, we develop a method that isolates eddies, fronts, and filaments in noisy SWOT measurements. 
We analyze variations in the sea surface height across spatial scales and use these statistics to estimate the turbulence signal, both within the swaths covered by SWOT's measurements and across the gap around the satellite's ground track, where only low-resolution measurements are available. 
We test this method using a realistic North Atlantic simulation, with added noise at short distances designed to mimic SWOT's measurements. 
The results show our method removes noise contamination while preserving characteristic features of eddies, fronts, and filaments down to a spatial scale of around \qty{6}{\kilo\meter}. 
This enables the use of SWOT data to study the ocean circulation at these scales.

\section{Introduction}

\begin{figure}[p]
  \centering
  \includegraphics[width=1.0\textwidth]{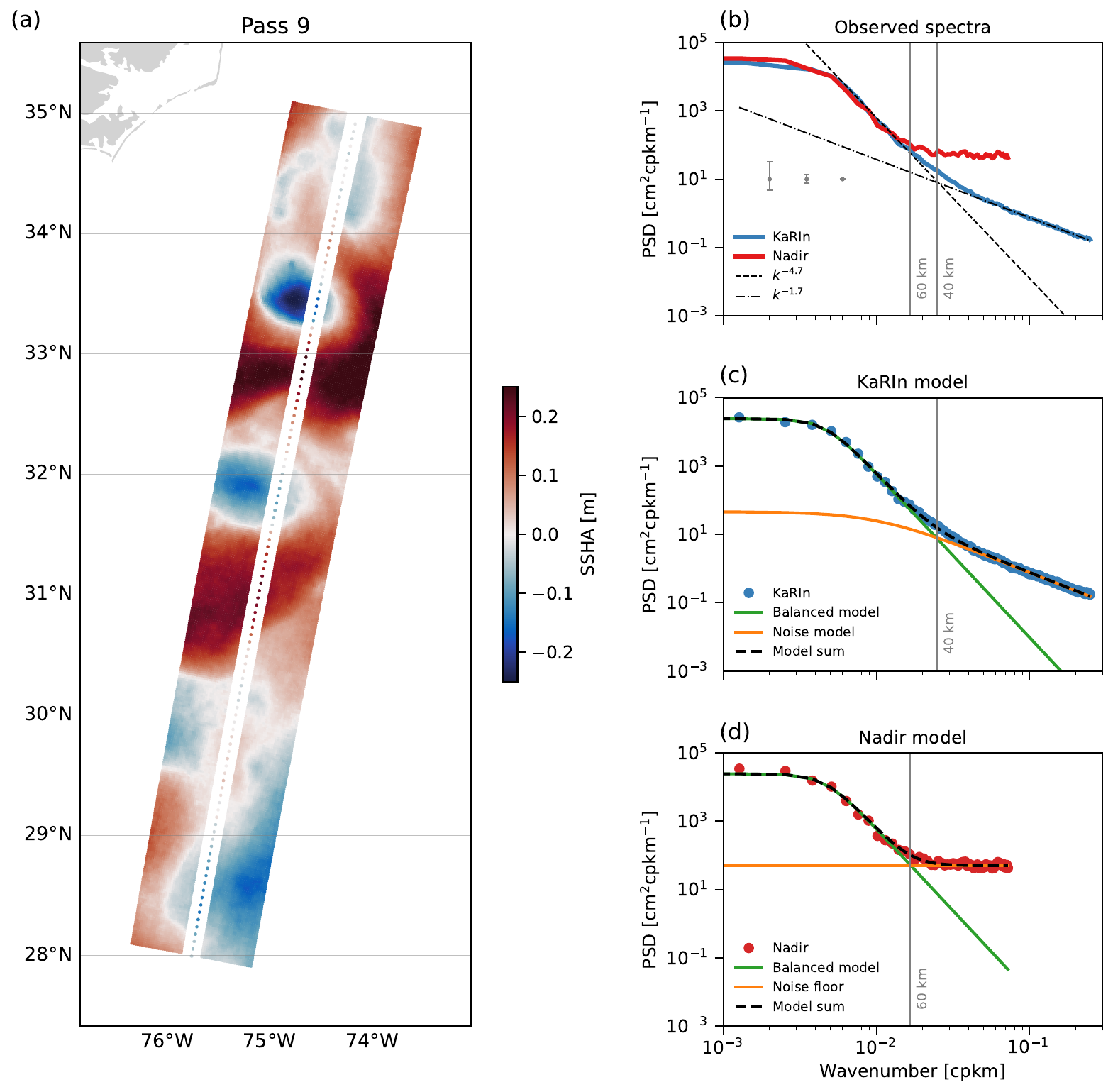}
  \caption{%
    SWOT sea surface height anomalies (SSHA) and their statistics in the Northwest Atlantic.
    (a)~SSHA from SWOT's KaRIn and nadir altimeters for pass~9, cycle~483 (7~April, 2023).
    (b)~Along-track SSHA variance spectra from the two altimeters, with reference lines 
    of $k^{-4.7}$ and $k^{-1.7}$ indicating the balanced and noise-dominated regimes in the KaRIn data.
    Gray error bars illustrate 95\% confidence intervals for the spectra at 1~degree of 
    freedom every ten cycles, 1~every cycle, and 1~every cycle and across-swath position (left to right, respectively).
    (c,~d)~Model fits to these spectra consisting of a balanced part and small-scale noise. 
    The sum of the models (black dashed lines) tracks the respective observed spectrum (dots).
    Vertical lines indicate the transition between the balanced and noise models: 
    $\qty{40}{\kilo\meter}$ for KaRIn and $\qty{60}{\kilo\meter}$ for the nadir altimeter.
  }
  \label{fig:swot_map_spectrum}
\end{figure}

The Surface Water and Ocean Topography (SWOT) mission is providing the first high-resolution two-dimensional maps of sea surface height (SSH), enabling a global observation of the ocean across a wide range of scales \parencite{Fuetal_2009, Morrowetal_2019, Morrow_etal_2018, Fuetal_2024}.
Unlike previous satellite altimetry missions, which measured SSH along one-dimensional nadir tracks with a \qtyrange{6}{12}{\kilo\meter} footprint, SWOT's wide-swath Ka-band Synthetic Aperture Radar Interferometer (KaRIn) instrument covers two \qty[number-unit-separator=-]{50}{\kilo\meter}-wide swaths, separated by a \qty{20}{\kilo\meter} nadir gap, the center of which is sampled by a traditional Jason-class altimeter \parencite[][see also Fig.~\ref{fig:swot_map_spectrum}a]{Neeck_etal_2012}.
Together, these measurements capture the SSH variability of the ocean from the mesoscale (\qtyrange{100}{300}{\kilo\meter}) down to part of the submesoscale (below \qty{100}{\kilo\meter}), providing the first two-dimensional altimeter observations of eddies, fronts, and filaments from space \parencite{Archer_2025}.

At the meso- and submesoscales, a large portion of the SSH variance arises from balanced flows.
Mesoscale eddies are nearly geostrophically balanced and are well described by quasi-geostrophic theory.
At the submesoscales, the assumption of geostrophic balance starts to break down \parencite[e.g.,][]{Thomas_etal_2008, Mcwilliams_2016, taylor&thompson_2023}.
This breakdown can have two flavors: balanced dynamics outside of the quasi-geostrophic regime, and unbalanced dynamics such as inertia--gravity waves and three-dimensional turbulence arising from unbalanced instabilities (gravitational, shear, symmetric, inertial, etc.).
To distinguish clearly between these two flavors of submesoscale motion, we use ``balanced flow'' to refer to any flow that can be described by theories that may go beyond the quasi-geostrophic regime and better capture ageostrophic effects but retain the idea that a dynamical balance can be used to infer the flow diagnostically from the distribution of potential vorticity \parencite[e.g.,][]{charney_use_1955,hoskins_geostrophic_1975,mcwilliams_intermediate_1980,muraki_next-order_1999}.
The balanced part of the flow is of particular interest because it often controls the transport of energy and tracers in the upper ocean.
Mesoscale eddies dominate the kinetic energy and control lateral stirring, while submesoscale balanced motions play a key role in restratifying the water column and regulating vertical exchange of physical and biogeochemical tracers \parencite[e.g.,][]{fox-kemper_parameterization_1_2008,Klein_2009,Mahadevan_2016,Mcwilliams_2016,levy_2012,levy_2018,taylor&thompson_2023}.
The balanced component of the circulation in the upper ocean leaves a signature in the surface height and can be inferred through the geostrophic and higher-order balance relations. 

Isolating the balanced part of the SSH signal in SWOT observations is a significant challenge, however, because the measured SSH is a complex superposition of the balanced signal along with a collection of other signals and measurement errors.  
\textcite{Zhang_2025} identified the following components in KaRIn data from typical mid-latitude open-ocean regions: balanced meso- and submesoscale signals, low-mode internal tides, a small-scale signal whose amplitude tracks the height of surface gravity waves, and geoid errors.
Geoid errors can be removed easily by subtracting the time-mean SSH, because the geoid does not vary in time. 
However, disentangling the remaining components requires more work and is the topic of this paper.
We focus on the Gulf Stream region, which has a strong balanced signal and comparatively weak tides, so low-mode internal tides do not contribute significantly to the signal. 
Other western boundary current regions and the Southern Ocean should also have these characteristics and be amenable to our analysis.
This focus on energetic regions leaves two dominant components.
The first is the SSH anomalies associated with the balanced circulation, which is the signal we aim to recover. The second is the small-scale signal, which we treat as ``noise.''
Despite the label, this ``noise'' may be a combination of measurement errors and real geophysical signals, such as those arising from the internal-wave continuum, and its spectral shape appears to be affected by the ground processing \parencite{Du_etal_2026}.
For the purpose of extracting the balanced signal, it is appropriate to lump these small-scale signals into a single noise term as long as its statistics are modeled appropriately.

These two components---the balanced signal and the small-scale noise---can be clearly identified in the wavenumber spectrum of SSH variance (Fig.~\ref{fig:swot_map_spectrum}b).
The balanced signal dominates at low wavenumbers, exhibiting a steep spectral drop-off 
(proportional to $k^{-4.7}$ in Fig.~\ref{fig:swot_map_spectrum}b) characteristic of balanced turbulence 
\parencite[e.g.,][and references therein]{Callies&Wu_2019}. 
The noise component dominates at high wavenumbers and has a much 
flatter spectrum (proportional to $k^{-1.7}$ in Fig.~\ref{fig:swot_map_spectrum}b).
The transition between the two occurs at an intermediate scale that depends on the relative amplitude of the two components ($\sim$\qty{40}{\kilo\meter} wavelength in Fig.~\ref{fig:swot_map_spectrum}b).
In the extraction described below, we assume that the power-law form of the balanced spectrum continues beneath the noise, effectively extrapolating the steep balanced slope into the smaller, unresolved scales.
This is consistent with available in~situ observations and high-resolution simulations that show SSH variance and kinetic-energy spectra typically maintain a single, continuous power law (with no spectral break) down to scales of a few kilometers \parencite[e.g.,][and Fig.~\ref{fig:synthetic_swot_data}b]{callies&ferarri_2013,callies_2015,Rocha_etal_2016,Callies&Wu_2019}.
We use this extrapolation as a Bayesian prior that can be revised should more information become available.

While one could devise a simple spatial filter that attempts to remove the small-scale noise, we pursue a more principled approach here that separates the observed signal into its balanced and noise components based on their distinct observed statistics.
As a choice of prior, we model both components as Gaussian processes with corresponding covariances to perform a Bayesian inversion that, given the data, infers the most probable balanced signal, along with an estimate of the uncertainty.
Our method jointly uses the KaRIn and the nadir data, fills the nadir gap, provides reliable error estimates, and can be extended to additional signals and more comprehensive statistical models.
The approach is based on fitting parametric models of the balanced signal and noise to the wavenumber spectrum of SSH variance for each instrument (Fig.~\ref{fig:swot_map_spectrum}b).
We evaluate the performance of this extraction of the balanced signal using synthetic SWOT observations generated from a high-resolution simulation of the subtropical North Atlantic.

\section{Data Sources and Processing} \label{sec:data}

Our analysis focuses on the SWOT mission's rapid-repeat phase from 26~March to 10~July 2023, during which the satellite had a $\qty{24}{\hour}$ repeat orbit.
We use two Level-2 data products: the Level~2 KaRIn Low Rate Sea Surface Height Data Product, Version C, which provides the swath measurements at \qty{2}{\kilo\meter} spatial resolution, and the Level~2 Nadir Altimeter Geophysical Data Record (GDR) with Waveforms, which samples SSH along the ground track with \qty{6.8}{\kilo\meter} resolution.
We use the sea surface height anomaly (SSHA) fields from both products (\texttt{ssha\_karin\_2} and \texttt{ssha}), which describe the deviation from the mean sea surface with external
and internal ocean tides, solid earth tides, load tides, pole tides, and the dynamic atmospheric correction removed.
Combined, these two datasets cover a \qty[number-unit-separator=-]{120}{\kilo\meter}-wide swath: KaRIn measures two \qty[number-unit-separator=-]{50}{\kilo\meter}-wide swaths on either side of the ground track, separated by a \qty{20}{\kilo\meter} gap, with the nadir altimeter sampling along the ground track in the center of the gap \parencite[Fig.~\ref{fig:swot_map_spectrum}a;][]{KaRIn,Nadir}.

\begin{figure}[t]
  \centering
  \includegraphics[width=0.7\textwidth]{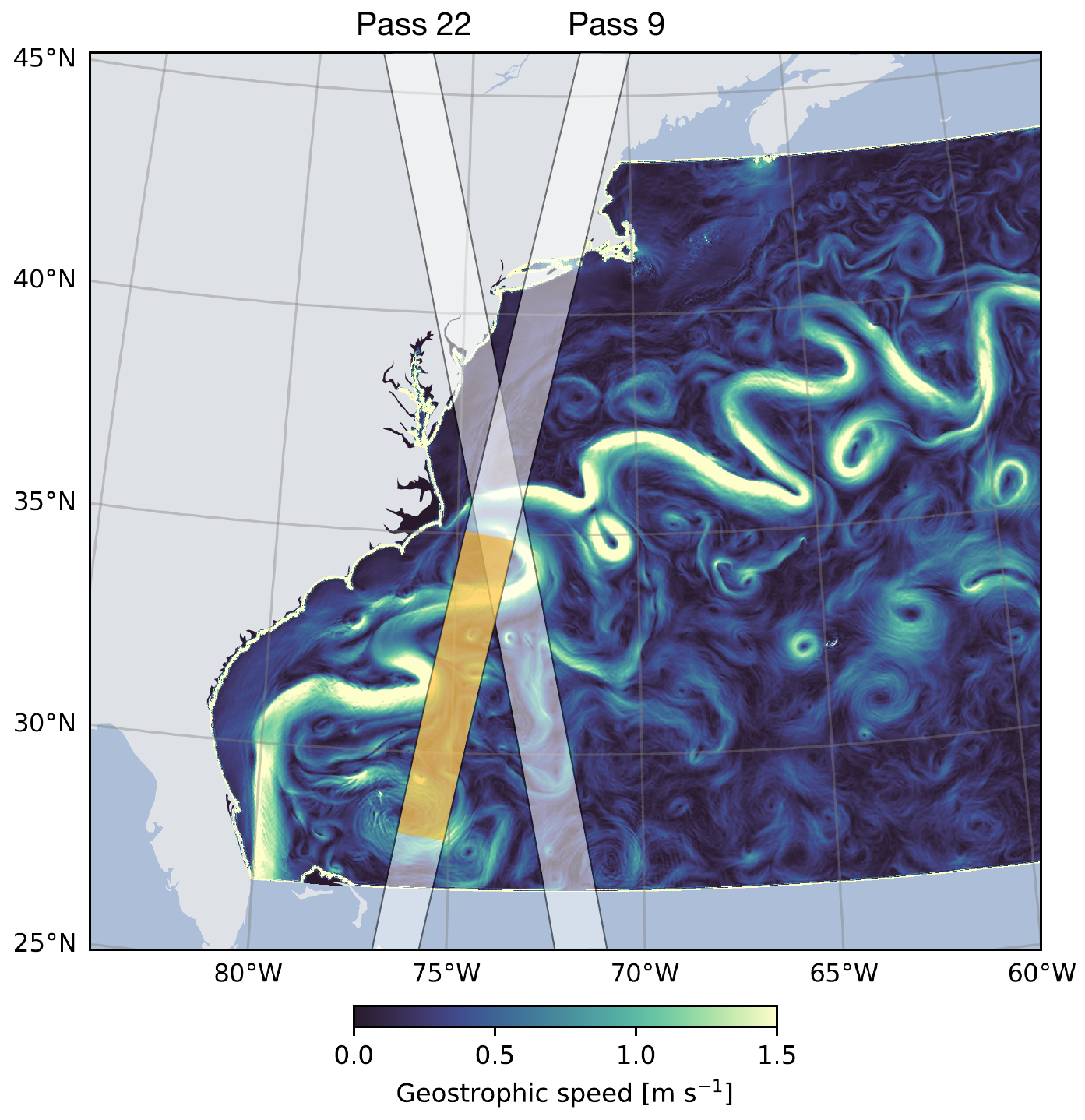}
  \caption{%
    Location of the SWOT rapid-repeat sampling in the Northwest Atlantic, overlaid on a map of surface geostrophic speed from the numerical simulation.
    The section of pass~9 used in the analysis is between \ang{28} and \ang{35}N and is highlighted in orange.
  }
  \label{fig:na_swot_passes}
\end{figure}

The statistical model we use for the balanced extraction method described below is widely applicable in regions with energetic balanced flows and comparatively weak internal tides. 
We illustrate and evaluate our method in the western subtropical North Atlantic using pass~9 of SWOT's rapid-repeat phase (Fig.~\ref{fig:na_swot_passes}).
We consider the section of the pass for which the ground track falls between \ang{28} and \ang{35}N.
This region offers a particularly clean test case: the balanced mesoscale and submesoscale flows are strong, and the SWOT observations for this pass contain minimal contamination (e.g., rain cells or other measurement artifacts).
In addition, we have ready access to a high-resolution simulation covering this region, which we will use below to test our method. 

We perform several preprocessing steps to prepare the SWOT data for analysis.
We apply two corrections: we add the crossover calibration height correction (\texttt{height\_cor\_xover}), and we restore the full internal-tide signal---part of which is subtracted by default in the distributed products---by adding back the HRET tidal correction (\texttt{internal\_tide\_hret}) provided with each dataset.
Although the HRET contribution is small in our focus region (at most $0.07\%$ of the SSHA power spectral density), we include it to avoid unintentionally removing any balanced low-wavenumber variance and to maintain consistency with our simulation data, where the full internal tide is retained.
Next, observation points from both the KaRIn and nadir instruments are removed if they exhibit non-zero quality flags associated with the SSHA measurement, rain, or ice.
For KaRIn, the omitted-point mask is dilated by 2~pixels in each direction to remove the neighbors of quality-flagged points. 
This accounts for imperfect quality flagging in the data product. 
Visual inspection of the SSH fields reveals that bad data (e.g., from precipitation cells) tend to cluster, with unflagged pixels adjacent to flagged ones often still containing errors.
We also remove any KaRIn cycle whose spatially averaged SSH variance exceeds ten times the variance computed across all cycles.
For both KaRIn and nadir, cycles in which more than 10\% of observation points are quality flagged are dropped entirely. 
We also remove the nadir observations of cycle~545 because of its anomalously high noise floor.
These steps remove 10~KaRIn cycles and 6~nadir cycles, meaning we retain 90\% and 94\% of the 101~available cycles for this pass, respectively; 87~of the cycles pass the quality control for both instruments.

Although the SWOT products already subtract a mean sea surface model, the mean sea surface variations due to small-scale topography were poorly constrained prior to launch \parencite{yu_abyssal_2024}.
As a result, the time-mean SSH field contributes substantially to the signal at \qtyrange{10}{50}{\kilo\meter} wavelength, producing a spectral bump \parencite{Zhang_2025}.
The rapid-repeat phase provides $\sim$\num{90} revisits per pass over a three-month period, yielding a robust estimate of the time mean at small scales.
At mesoscales, this time window is too short to average out real ocean variability, meaning the time-mean field retains signals from transient mesoscale eddies that should not be subtracted.
This is not a problem, however, because the geoid is already well-constrained for scales above \qty{100}{\kilo\meter} \parencite[e.g.,][]{Andersen_2012, Andersen_2023}.
We therefore apply a Gaussian high-pass filter with a \qty{100}{\kilo\meter} full width at half maximum (FWHM) scale to the time mean, isolating the small-scale correction, and subtract the resulting filtered time-mean field from each KaRIn swath. 
We also remove the spatial mean, calculated from the KaRIn data and subtracted from both the KaRIn and the nadir data for each cycle, to eliminate signals associated with large-scale mixed-layer heating and cooling \parencite{gill_theory_1973}.

We estimate one-dimensional along-track wavenumber spectra of SSHA variance for both instruments (Fig.~\ref{fig:swot_map_spectrum}b).
The spectra are computed from SWOT ground tracks within the latitude range \ang{28} to \ang{35}N, corresponding to approximately \qty[number-unit-separator=-]{790}{\kilo\meter}-long segments, which provides sufficient coverage of the mesoscale band.
To reduce spectral leakage, each segment is tapered with a sine-squared (Hann) window, 
normalized to preserve the variance of a stationary signal, before a discrete Fourier transform is applied in the along-track direction.
For the KaRIn data, the power spectral density (PSD) is then estimated by averaging the resulting spectrograms across the track (for both swaths) and over all cycles.
Any across-track positions containing masked values are excluded before this averaging.
For the nadir data, the spectrograms are averaged over all available cycles from the rapid-repeat phase.

Throughout this paper, spectra are shown as functions of ordinary wavenumber~$k$ (units of \si{\cpkm}), the inverse of the wavelength $\ell$.
The angular wavenumber $2\pi k$ and its inverse are more easily associated with the scales of features; e.g., the Fourier transform of a Gaussian of width~$\sigma$ is a Gaussian of width $\sigma^{-1}$ in angular wavenumber space and $(2\pi \sigma)^{-1}$ in ordinary wavenumber space.
This conversion factor should be kept in mind when interpreting the ordinary wavenumber spectra throughout the paper.
Furthermore, all spectra are one-sided, i.e., normalized such that  
\begin{equation}
    \langle h^2 \rangle = \int_0^\infty P(k)\, \mathrm{d} k,
\end{equation}
where $h$ is the SSHA, angle brackets denote the average, and $P(k)$ is the PSD.

To test the balanced extraction method, we generate synthetic SWOT data using a regional simulation of the midlatitude North Atlantic.
The simulation is performed at (1/48)\textdegree{} horizontal resolution ($\sim$\qty{2.3}{\kilo\meter}), uses a realistic but annually repeated atmospheric forcing, and has realistic tides forced through the open boundary conditions and the astronomical gravitational potential \parencite{Sinha_2023, SkinnerLawrence&Callies_2025}.
The simulation domain spans the width of the North Atlantic, but our analysis focuses on the subtropical Northwest Atlantic, where SWOT passes 9 and~22 intersect the Gulf Stream (Fig.~\ref{fig:na_swot_passes}), and strong balanced eddies and fronts are present.
We sample the simulated SSH output as hourly snapshots, at the spatial coordinates of SWOT pass~9 for all cycles that pass our quality control.
For each KaRIn and nadir measurement, we match its observation time stamp to the corresponding day of year and hour in the simulation output and then use two-dimensional cubic splines to spatially interpolate the corresponding simulation field onto the observation locations.

We subtract a long-term time mean from the simulation data to remove the mean sea surface field.
Following the SWOT processing, we also subtract a spatial mean for each cycle, computed over the region sampled by the KaRIn swaths.
In the SWOT product, the barotropic tide and the inverted barometer (IB) signal are removed separately using model-based corrections.
Because both signals are dominated by scales much larger than the swath, subtracting the spatial mean effectively removes them from the simulation output as well.
Any residual barotropic tide or IB signal has no effect on our results because its SSHA variance is orders of magnitude smaller than that of the energetic balanced SSHA field.

We use the simulation as a test bed for the balanced extraction, treating its underlying geophysical fields as known and providing a ``ground truth.''
Because the simulation does not assimilate any data and uses a repeat year for the forcing, its atmospheric and oceanic states differ from those of the real system during the SWOT observations.
This does not affect our ability to assess the extraction method, however, because we view both the observations and the simulation as providing draws from a statistical distribution of a chaotic system.
We only require that the simulation's statistics be sufficiently realistic, which we evaluate below (see Fig.~\ref{fig:synthetic_swot_data}b), not that the simulated fields match the observed fields exactly.

\section{Extraction of Balanced Signals}

To extract the balanced signal from the SWOT data, we adopt a parametric framework similar to that of \textcite{Zhang_2025}.
We extend their method in several ways to fully exploit the SWOT observations.
First, we fit spectral models jointly to both the KaRIn and nadir measurements, enabling a consistent treatment of the complete SWOT dataset.
Second, rather than prescribing the slope of the balanced spectrum, we treat it as a free parameter to be estimated directly from the data, allowing the model to adapt more finely to the observed statistics.
Third, we use the resulting covariances in a Bayesian inversion to reconstruct the balanced field across the entire swath, including the nadir gap.
This framework also interpolates over masked or missing data points, using the inferred spatial covariances to provide statistically consistent estimates of the balanced field in regions without reliable observations.
Together, these extensions provide a more flexible and data-driven characterization of the balanced signal and enable a comprehensive estimation of the balanced SSHA field.

\subsection{Parametric Models for Balanced and Noise Spectra}
\label{sec:spectral_models}

We assume that the SSHA observations are the sum of a balanced signal and noise.
The two instruments have distinct noise characteristics but sample the same balanced signal (Fig.~\ref{fig:swot_map_spectrum}b).
The nadir altimeter has an approximately white noise spectrum, while the \qty{2}{\kilo\meter} KaRIn signal has a red noise spectrum of substantially reduced amplitude.
As discussed in the introduction, this part of the KaRIn signal is incompletely understood but appears to be associated with surface gravity waves \parencite{Zhang_2025}. 
It is substantially affected by the ground processing; specifically, the smoothing applied to the \qty{250}{\meter} data to produce the \qty{2}{\kilo\meter} product can turn a white noise floor into a red along-track spectrum \parencite{Du_etal_2026}.
While real (unbalanced) signals such as the internal-wave field may also contribute to this part of the spectrum, we summarily refer to it as ``noise'' that obscures our balanced target signal.

As in \textcite{Zhang_2025}, we adopt the following parametric models for the one-dimensional spectrum of the balanced signal and KaRIn's noise:
\begin{equation}
  B(k) = \frac{A_{\mathrm{b}}}{1 + (\lambda_{\mathrm{b}} k)^{\, s_{\mathrm{b}}}}, \qquad N(k) = \frac{A_{\mathrm{n}}}{\left[1 + (\lambda_{\mathrm{n}} k)^2\right]^\frac{s_{\mathrm{n}}}{2}}.
  \label{eq:balanced_model}
\end{equation}
Here, $A_\mathrm{b}$ and $A_\mathrm{n}$~set the amplitude of the spectra, $\lambda_\mathrm{b}$~and $\lambda_\mathrm{n}$ are the wavelengths at which the spectra transition from a low-wavenumber plateau to a high-wavenumber power law, and $s_\mathrm{b}$ and $s_\mathrm{n}$ are the exponents of those power laws.
By treating $s_\mathrm{b}$ and $s_\mathrm{n}$ as free parameters, we allow the SWOT data to determine the field's local smoothness rather than prescribing it a priori \parencite[e.g.,][]{Stein_1999,rasmussen_gaussian_2005}.
The power-law behavior at high wavenumbers offers a much better statistical model than the often-used squared exponential covariance function \parencite[e.g.,][]{bretherton_technique_1976}.
We thus model the observed KaRIn and nadir spectra as
\begin{equation}
P^\mathrm{K}(k) = B(k) + N(k),
\label{eqn:karin_model}
\end{equation}
\begin{equation}
P^\mathrm{N}(k) = B(k) + 2\Delta \sigma^2,
\label{eqn:nadir_model}
\end{equation}
where the white noise term in the nadir model derives its magnitude from the standard deviation~$\sigma$ and spacing~$\Delta$ of the nadir measurements.
We assume throughout this paper that the balanced signal is not correlated to the noise of either instrument and that the noise of the two instruments is mutually uncorrelated.
The latter assumption is a reasonable starting point, given the instruments' distinct measurement physics and noise characteristics, but it could be revised in the future when KaRIn's noise is better understood.
Furthermore, even if geophysical signals contribute to the small-scale signal in KaRIn data, it remains appropriate to lump them in with measurement errors and assign them to the KaRIn instrument. If such geophysical signals contribute to the nadir data, they are overwhelmed by the much larger instrument noise and need not be modeled.
We also assume that all statistics are isotropic and stationary in space and time.
We discuss the latter assumption in Section~\ref{sec:discussion} and describe ways in which it could be relaxed.

We estimate the parameters $A_\mathrm{b}$, $\lambda_\mathrm{b}$, $s_\mathrm{b}$, $A_\mathrm{n}$, $s_\mathrm{n}$, and $\sigma$ from the observations by fitting the models~\eqref{eqn:karin_model} and~\eqref{eqn:nadir_model} to the PSD estimated from the data.
We set $\lambda_\mathrm{n} = \qty{100}{\kilo\meter}$ because it is poorly constrained by the data and has little influence on the model fits, provided it is large enough.
The remaining parameters for $B(k)$ and $N(k)$ are obtained by fitting the KaRIn spectrum, and $\sigma$ is obtained by fitting the nadir spectrum while holding the parameters for $B(k)$ fixed, because the KaRIn data provide a much stronger constraint on them than the nadir measurements.
All fits are performed in log--log space using a least-squares method with a $k^{-1}$ weighting factor following \textcite{lawrence_2022,Zhang_2025}.
This prevents the densely sampled high-wavenumber data from dominating the fit, ensuring a stable fit across all scales.

Applying this procedure to pass~9 yields good fits to the spectra estimated from the SWOT observations (Fig.~\ref{fig:swot_map_spectrum}c,d).
The fit produces a balanced spectrum that transitions from a plateau at $A_\mathrm{b} = \qty{2.4e4}{\centi\meter\squared\per\cpkm}$ to a power law at a wavelength of $\lambda_\mathrm{b} = \qty{216}{\kilo\meter}$, and it estimates the spectral slope to be $s_\mathrm{b} = 4.7$.
Dividing the transition scale~$\lambda_\mathrm{b}$, which corresponds to the size of the dominant mesoscale eddies, by $2\pi$ corresponds to a spatial scale of $\sim$\qty{40}{\kilo\meter}, consistent with the radius of the strongest eddies in the maps (Fig.~\ref{fig:swot_map_spectrum}a).
The noise has a much lower plateau at $A_\mathrm{n} = \qty{45.6}{\centi\meter\squared\per\cpkm}$ and a smaller spectral slope of $s_\mathrm{n} = 1.7$.
The nadir noise is estimated to have an amplitude $\sigma = \qty{1.9}{\centi\meter}$.
The balanced and noise models intersect at a wavelength of $\qty{40}{\kilo\meter}$ for the KaRIn data and $\qty{60}{\kilo\meter}$ for the nadir data.

\subsection{Estimating the Balanced Signal}
\label{sec:GP}

With the parameters of the spectral models in hand, we perform a statistical estimate of the balanced component of the SSHA field.
We adopt a Gaussian process framework to combine the observations from the KaRIn and nadir instruments.
The observations, denoted $\vec{h}_\mathrm{K}$ for KaRIn and $\vec{h}_\mathrm{N}$ for the nadir altimeter, are modeled as a combination of a balanced signal at their respective locations, $\xk$ for KaRIn and $\xn$ for nadir, plus an instrument-dependent noise.
Our goal is to estimate the balanced signal $\vec{h}_*$ at a set of target locations $\vec{x}_*$.
We assume that both the signal and noise components have Gaussian statistics.
Under this assumption, the joint distribution of the target balanced signal~$\vec{h}_*$ and the observations~$\vec{h}_\mathrm{K}$ and $\vec{h}_\mathrm{N}$ is a multivariate Gaussian.
This formulation is well suited for capturing the second-order statistics (i.e., the covariance and spectrum), although higher-order moments may exhibit non-Gaussian characteristics (see the discussion below).
The joint Gaussian prior distribution is given by
\begin{equation}
  \begin{bmatrix}
    \vec{h}_\mathrm{K} \\
    \vec{h}_\mathrm{N} \\
    \vec{h}_*
  \end{bmatrix}
  \sim \mathcal{N} \left( \vec{0},
    \begin{bmatrix}
      \mat{R}_\mathrm{KK} & \mat{R}_\mathrm{NK}\T & \mat{R}_\mathrm{*K}\T \\
      \mat{R}_\mathrm{NK} & \mat{R}_\mathrm{NN} & \mat{R}_\mathrm{*N}\T \\
      \mat{R}_\mathrm{*K} & \mat{R}_\mathrm{*N} & \mat{R}_\mathrm{**}
    \end{bmatrix}
  \right).
\end{equation}
The covariance matrices are obtained from the spectral models above using the Wiener--Khinchin theorem, assuming stationarity and isotropy of the statistics.
With the cosine transform written as
\begin{equation}
  \mathcal{C}[P](r) \equiv \int_0^\infty P(k) \cos(2\pi k r) \, \d k,
\end{equation}
the observational covariances are
\begin{align}
  (R_\mathrm{NN})_{ij} &= \mathcal{C}[B(k)](r_{ij}) + \sigma^2 \delta_{ij}, \\
  (R_\mathrm{KK})_{ij} &= \mathcal{C}[B(k) + N(k)](r_{ij}), \\
  (R_\mathrm{NK})_{ij} &= \mathcal{C}[B(k)](r_{ij}),
\end{align}
and the covariances involving the target are
\begin{align}
  (R_\mathrm{**})_{ij} &= \mathcal{C}[B(k)](r_{ij}), \\
  (R_\mathrm{*N})_{ij} &= \mathcal{C}[B(k)](r_{ij}), \\
  (R_\mathrm{*K})_{ij} &= \mathcal{C}[B(k)](r_{ij}),
\end{align}
where $r_{ij}$ denotes the distance between the points under consideration. 
Note that the noise only appears in the KaRIn--KaRIn and nadir--nadir covariances.
We evaluate the cosine transforms numerically (see Appendix~\ref{sec:smoothing}), which allows for flexible covariance functions adapted to the observed spectral slopes.

With this prior, the posterior distribution of the balanced SSHA field $\vec{h}_*$ conditioned on the observations follows from Bayes's theorem \parencite[e.g.,][]{sanz-alonso_inverse_2023}:
\begin{equation}
  \vec{h}_* \, | \, \vec{h}_\mathrm{K}, \vec{h}_\mathrm{N} \sim \mathcal{N}(\vec{m}, \mat{C}).
\end{equation}
The posterior mean $\vec{m}$ provides the best estimate of the balanced signal, and the posterior covariance $\mat{C}$ quantifies its uncertainty.
They are given by
\begin{equation}
  \vec{m} =
  \begin{bmatrix}
    \mat{R}_\mathrm{*K} & \mat{R}_\mathrm{*N}
  \end{bmatrix}
  \begin{bmatrix}
    \mat{R}_\mathrm{KK} & \mat{R}_\mathrm{NK}\T \\
    \mat{R}_\mathrm{NK} & \mat{R}_\mathrm{NN}
  \end{bmatrix}^{-1}
  \begin{bmatrix}
    \vec{h}_\mathrm{K} \\
    \vec{h}_\mathrm{N}
  \end{bmatrix} \label{eq:post_mean}
\end{equation}
and
\begin{equation} \label{eq:cov}
  \mat{C} = \mat{R}_{**} -
  \begin{bmatrix}
    \mat{R}_\mathrm{*K} & \mat{R}_\mathrm{*N}
  \end{bmatrix}
  \begin{bmatrix}
    \mat{R}_\mathrm{KK} & \mat{R}_\mathrm{NK}\T \\
    \mat{R}_\mathrm{NK} & \mat{R}_\mathrm{NN}
  \end{bmatrix}^{-1}
  \begin{bmatrix}
    \mat{R}_\mathrm{*K} & \mat{R}_\mathrm{*N}
  \end{bmatrix}\T.
\end{equation} 
We perform the matrix inversion using a Cholesky decomposition.

In the above, we consider all parameters of the spectral models as fixed by the fits, and we do not propagate any uncertainty arising from the spectral estimation uncertainty.
Given that many degrees of freedom are available at the crucial scales at which the spectrum transitions from the balanced-dominated signal to the noise-dominated signal, the spectral estimation uncertainty is small (Fig.~\ref{fig:swot_map_spectrum}b) and likely negligible in its impact on the posterior.

\begin{figure}[p]
  \centering
  \includegraphics[width=\textwidth]{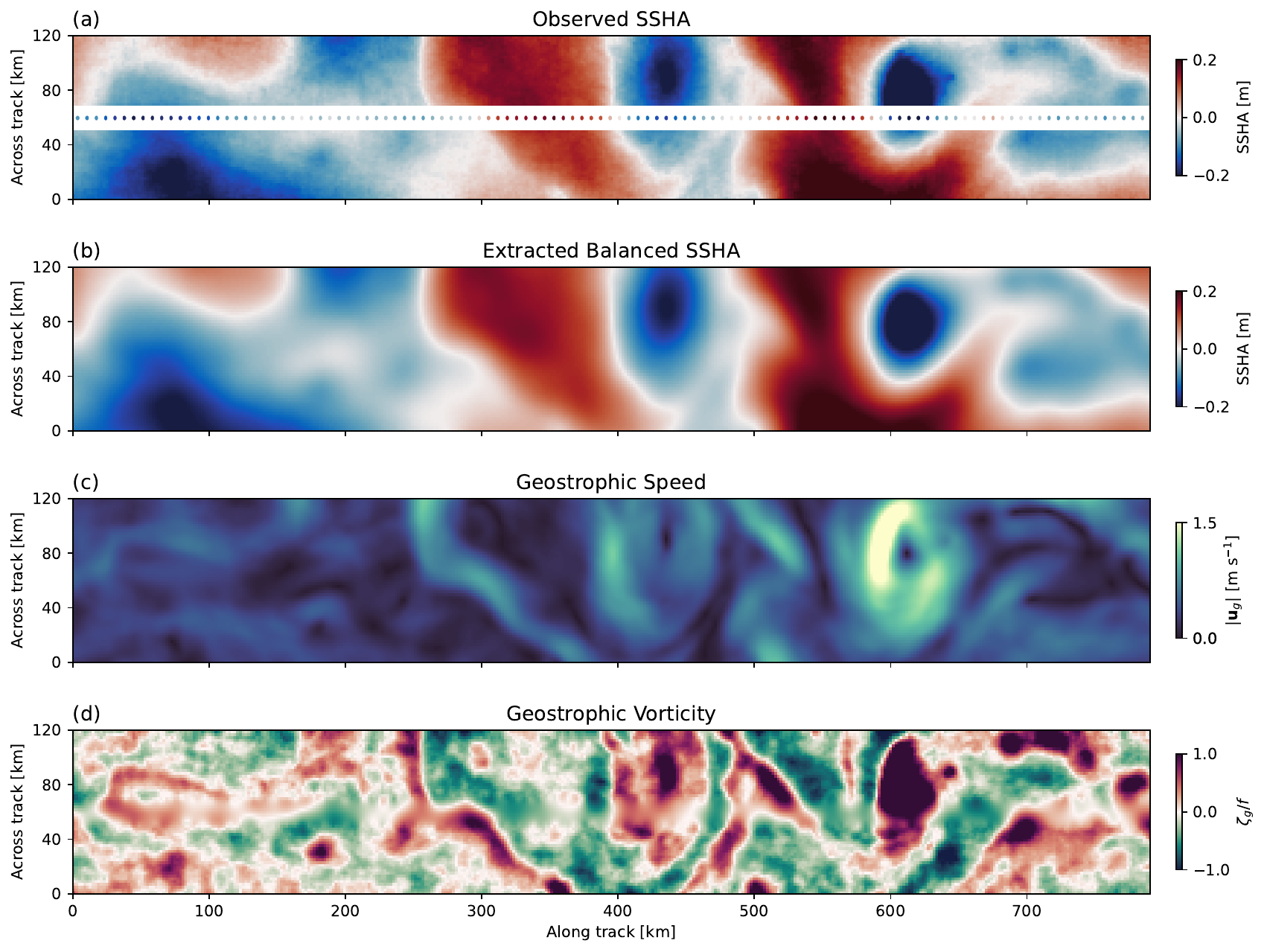}
  \caption{%
    Balanced extraction applied to the observed SWOT SSHA field on pass~9, cycle~483 (7~April, 2023).
    (a)~Observed SSHA data from KaRIn (swaths) and nadir (central track).
    (b)~Extracted balanced SSHA field estimated from the observed SSHA.
    (c)~Geostrophic speed~$|\vec{u}_g|$ of the extracted balanced SSHA field.
    (d)~Geostrophic vorticity~$\zeta_g$ derived from the extracted balanced SSHA and normalized by the planetary vorticity~$f$.
    The orientation of these maps is such that the lower left corner corresponds to the southeast corner of the patch shown in Fig.~\ref{fig:na_swot_passes}.
  }
  \label{fig:balanced_extraction}
\end{figure}

This estimation of the balanced signal~\eqref{eq:post_mean}, from here on referred to as the balanced extraction, effectively separates balanced meso- and submesoscale structures from the high-wavenumber noise, as illustrated using a sample of the energetic Gulf Stream region on 7~April 2023.
The SSHA data from SWOT reveal a pair of mesoscale cyclones, approximately \qtyrange{50}{75}{\kilo\meter} in diameter, as well as sharp frontal features between them and in their surroundings (Fig.~\ref{fig:balanced_extraction}a).
One front, corresponding to a sharp gradient in SSHA, extends across the entire width of the swath and from $\sim$\qtyrange{250}{350}{\kilo\meter} in the along-track direction.
Another front can be seen between the eddies at an along-track position of $\sim$\qty{500}{\kilo\meter}.
These structures are bisected by the \qty{20}{\kilo\meter} nadir gap, with small-scale noise superimposed on them.

When applied to these observations, the balanced extraction yields a reconstruction that preserves the mesoscale eddies and frontal structures while suppressing small-scale noise and seamlessly bridging the nadir gap (Fig.~\ref{fig:balanced_extraction}b).
The posterior mean field $\vec{m}$, reshaped into a map for visualization, closely resembles the original SSHA but is noticeably smoother due to the removal of the small-scale noise.
The extracted balanced signal retains the sharp SSHA drops across the aforementioned fronts, although the gradients are smoother.
Multiple small-scale eddies with a diameter of $\sim$\qty{10}{\kilo\meter} can be seen, e.g., located around $(x, y) = (\qty{180}{\kilo\meter}, \, \qty{30}{\kilo\meter})$, $(x, y) = (\qty{500}{\kilo\meter}, \, \qty{0}{\kilo\meter})$, and $(x, y) = (\qty{250}{\kilo\meter}, \, \qty{40}{\kilo\meter})$, where $x$ is the along-track coordinate and $y$ is the across-track coordinate (Fig.~\ref{fig:balanced_extraction}a,b).
We quantify the smallest features that the balanced extraction can reconstruct below.

The dynamical structures in the balanced SSHA field are clarified by examining its derivatives, which highlight small-scale features that are less apparent in the surface height field (Fig.~\ref{fig:balanced_extraction}c,d).
We calculate the geostrophic velocity components (in the swath-aligned coordinates) and the geostrophic speed,
\begin{equation}
  u_g = -\frac{g}{f} \frac{\partial h}{\partial y}, \qquad
  v_g = \frac{g}{f} \frac{\partial h}{\partial x}, \qquad
  |\vec{u}_g| = \sqrt{u_g^2 + v_g^2},
\end{equation}
as well as the geostrophic vorticity
\begin{equation}
  \zeta_g = \frac{\partial v_g}{\partial x} - \frac{\partial u_g}{\partial y} = \frac{g}{f} \nabla^2 h,
\end{equation}
where $g$~denotes the gravitational acceleration and $f$ the planetary vorticity.
To compute the geostrophic velocities and vorticity from the balanced SSHA field, we use standard second-order centered finite-difference stencils in the interior and corresponding one-sided second-order stencils at the boundaries. 
We have checked these calculations against fourth-order accurate stencils (5-point centered in the interior and corresponding one-sided stencils of the same order at the boundaries) and found that both approaches produce indistinguishable results in the interior, reflecting the smoothness of the extracted balanced SSHA fields. 
The higher-order stencils introduce slightly more noise near the boundaries, so we retain the second-order formulation.

The resulting map of geostrophic speed shows regions of energetic flow around the peripheries of the eddies and along the fronts (Fig.~\ref{fig:balanced_extraction}c).
The cyclone located at $\sim$\qty{600}{\kilo\meter} along-track distance exhibits a peak geostrophic speed of \qty{2.1}{\meter\per\second}, although the actual speed is expected to be lower due to deviations from geostrophic balance (because $\zeta/f \gtrsim 1$, see Fig.~\ref{fig:balanced_extraction}d).
The corresponding geostrophic vorticity field clearly exhibits the two prominent mesoscale cyclones apparent in the SSHA field as well as a variety of much smaller $\sim$\qty{10}{\kilo\meter} cyclonic and anticyclonic flow features in the surroundings (Fig.~\ref{fig:balanced_extraction}d).
There are both elongated features along fronts and more circular submesoscale eddy features.
The cyclonic features are noticeably stronger and more localized than the anticyclonic features.
 
\begin{figure}[p]
  \centering
  \includegraphics[width=0.5\textwidth]{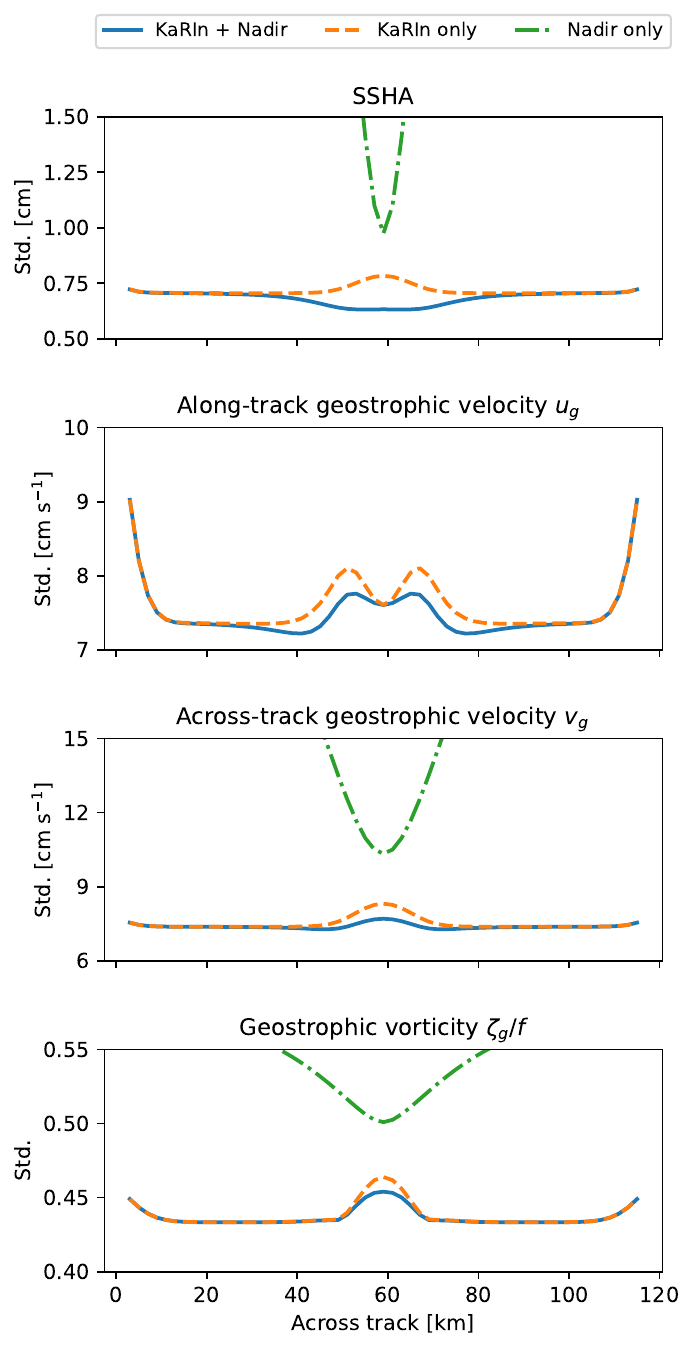}
  \caption{%
    Across-track profiles of the pointwise posterior uncertainty, derived from the posterior covariance matrix~$\mat{C}$.
    Shown are the standard deviations for SSHA, the along-track geostrophic velocity~$u_g$, the across-track geostrophic velocity~$v_g$, and the normalized geostrophic vorticity~$\zeta_g/f$. 
    Each are shown for the full extraction (KaRIn + Nadir; blue lines) and two reduced extractions in which either the KaRIn or nadir data are withheld (green and orange lines, respectively).
    The nadir measurements are located at \qty{60}{\kilo\meter} across-track distance. 
    }
  \label{fig:nadir_vs_nonadir}
\end{figure}

We quantify the uncertainty in the balanced extraction by evaluating the posterior covariance $\mat{C}$ from~\eqref{eq:cov}.
The square root of the diagonal of this matrix yields the standard deviation, or the expected pointwise error, of the extraction.
The spatial structure of this error reflects the sampling geometry and noise correlation, with the lowest uncertainty within the nadir gap and the highest uncertainty at the KaRIn swath edges (Fig.~\ref{fig:nadir_vs_nonadir}a).
In the nadir gap, the estimate draws on both the KaRIn and nadir data, and because their errors are assumed to be uncorrelated, the combined uncertainty is lower than for an estimate using either instrument individually.
Within the swath, the estimate also draws on both sources, but the contribution of the nadir data diminishes with distance from the ground track, so the uncertainty rises toward the KaRIn-only level.
In the centers of the KaRIn swaths (\qty{30}{\kilo\meter} and \qty{90}{\kilo\meter}), the pointwise uncertainty is about \qty{0.70}{\centi\meter}, and it falls to \qty{0.64}{\centi\meter} at the nadir (\qty{60}{\kilo\meter}).
If we withhold the nadir data, the uncertainty increases to \qty{0.78}{\centi\meter} at the nadir and becomes nearly flat at \qty{0.70}{\centi\meter} within the KaRIn swaths.
If we withhold the KaRIn data, it increases to \qty{1.0}{\centi\meter} at the nadir.

The error can easily be propagated to the geostrophic velocity components and the geostrophic vorticity, because these are all linear functions of the SSHA estimate.
Given a matrix $\mat{L}$ that maps the estimate~$\vec{m}$ of $h$ to the estimate $\vec{y} = \mat{L} \vec{m}$ of $u_g$, $v_g$, or $\zeta_g$, constructed as a finite-difference matrix multiplied by the appropriate constant, the error covariance of the derived field is $\mat{L} \mat{C} \mat{L}^\mathrm{T}$.
This uncertainty propagation takes full account of the spatial error correlations and of how these propagate through the finite differencing into the derived fields.
In the centers of the KaRIn swaths, the uncertainty for the two velocity components is \qty{7.3}{\centi\meter\per\second} and increases slightly in the nadir gap (Fig.~\ref{fig:nadir_vs_nonadir}b,c).
The uncertainty of the along-track velocity~$u_g$ has a double-peaked structure in the nadir gap.
This structure is due to KaRIn's sampling geometry and is present also if the data from the nadir altimeter are withheld.
The reason for the local uncertainty minimum at the nadir is that the across-track SSHA gradient can be estimated more accurately in the center of the nadir gap than off the center.
The gradient is better constrained when the inner edges of the KaRIn swaths are both \qty{10}{\kilo\meter} away than when one of the swaths is closer and one is farther off.
The uncertainty of the across-track velocity~$v_g$ has a single peak at the nadir because that is where the KaRIn data constrain the along-track gradient least.
The nadir data somewhat reduce the uncertainties of $u_g$ and $v_g$ near the local uncertainty maxima.
By itself, the nadir altimeter can only constrain the along-track gradient; the nadir-only estimate of~$v_g$ has an uncertainty of \qty{10}{\centi\meter\per\second} at the nadir, substantially larger than even the maximum uncertainty of the KaRIn-only estimate at the nadir.
The uncertainty of~$\zeta_g/f$ is \num{.433} in the centers of the KaRIn swaths and rises to \num{.454} at the nadir (Fig.~\ref{fig:nadir_vs_nonadir}d).
If the nadir data are withheld, the uncertainty increases to \num{.464} at the nadir but remains virtually the same within the KaRIn swaths.
The nadir altimeter by itself produces an uncertainty that is \qty{.50} at the nadir and rises toward the prior uncertainty of \num{.575} away from the ground track.

These posterior uncertainties indicate that many of the reconstructed flow features are statistically significant.
For SSHA, the reconstruction captures mesoscale eddies with amplitudes reaching \qty{+-30}{\centi\meter}, well above the uncertainty of less than \qty{1}{\centi\meter}.
Similarly, the balanced geostrophic velocity field resolves sharp geostrophic currents of up to \qty{2.1}{\meter\per\second}, greatly exceeding the in-swath uncertainty of \qty{7}{\centi\meter\per\second}.
For geostrophic vorticity, despite the uncertainty being a substantial fraction of~$f$, many flow features rise out of this uncertainty, with values up to $3.7f$ in the strong eddy cores.
We again caution that the actual currents and vorticity of the surface flow will substantially differ from their geostrophic values where $|\zeta_g| \gtrsim f$. 
The increased relative uncertainty in the geostrophic vorticity reflects the increased importance of small-scale features, which are captured by the extraction less confidently because of the dominance of noise at small scales.
Nevertheless, there are many submesoscale fronts, filaments, and eddies whose geostrophic vorticity exceeds the uncertainty substantially.

\begin{figure}[t]
  \centering
  \includegraphics[width=\textwidth]{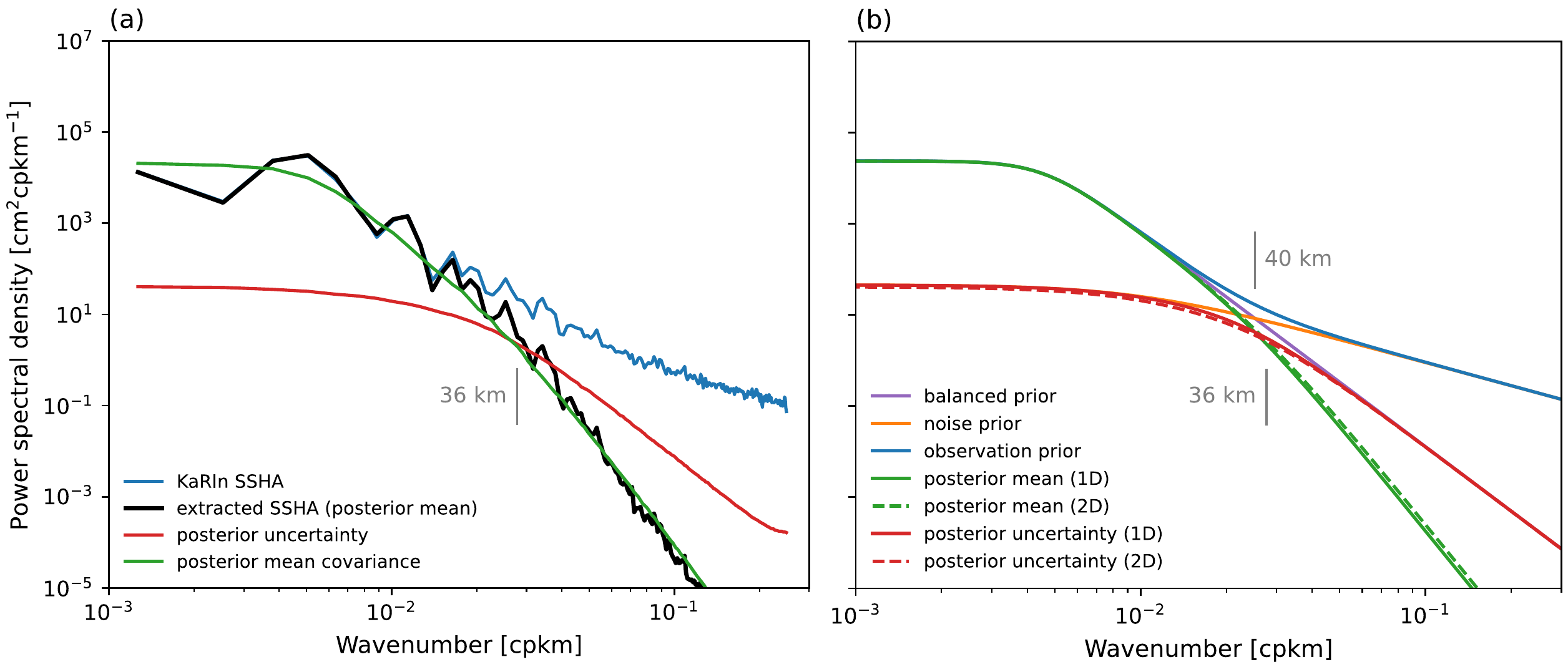}
  \caption{%
    (a) Along-track SSHA variance spectra for SWOT pass~9, cycle~483 (7~April, 2023).
    Shown are the spectra computed directly from KaRIn data (blue) and from the extracted balanced field (black).
    Also shown are the spectra computed from draws from the probability distributions of the posterior uncertainty (red) and the posterior mean~\eqref{eqn:postmeancov} (green).
    Their intersection at \qty{36}{\kilo\meter} marks the effective resolution limit of the extracted signal.
    (b) Fourier-space model of the balanced extraction. Priors for the balanced signal $B(k)$ (purple), noise $N(k)$ (orange), and observations (blue) are shown alongside 1D (solid) and 2D (dashed) posterior mean (green) and uncertainty (red). Accounting for the 2D swath geometry reduces the effective resolution limit from \qty{40}{\kilo\meter} (1D) to \qty{36}{\kilo\meter} (2D), matching panel~(a).
  }
  \label{fig:balanced_spectrum}
\end{figure}

The effective resolution of the balanced extraction can be assessed by comparing the power spectra of the posterior mean and the posterior uncertainty.
The spectra of the original KaRIn data and the extracted SSHA fields at the KaRIn sampling locations track each other at low wavenumbers, including the steep mesoscale-to-submesoscale slope (Fig.~\ref{fig:balanced_spectrum}a).
The spectrum of the extracted balanced signal diverges from the KaRIn observations and rolls off more steeply beyond $\sim$\qty{70}{\kilo\meter}, reflecting a strong attenuation of the extracted signal in the high-wavenumber range, where the original observation is dominated by noise.
To quantify where uncertainty begins to dominate, we draw random realizations from the probability distribution of the posterior uncertainty, characterized by the covariance~$\mat{C}$, by multiplying the Cholesky factor of~$\mat{C}$ with random vectors drawn from a standard normal distribution, and we compute the spectrum averaged over 200 realizations.
We likewise estimate the spectrum of the posterior mean by drawing from its probability distribution, characterized by the covariance
\begin{equation}
  \langle \vec{m} \vec{m}\T \rangle =
  \begin{bmatrix}
    \mat{R}_\mathrm{*K} & \mat{R}_\mathrm{*N}
  \end{bmatrix}
  \begin{bmatrix}
    \mat{R}_\mathrm{KK} & \mat{R}_\mathrm{NK}\T \\
    \mat{R}_\mathrm{NK} & \mat{R}_\mathrm{NN}
  \end{bmatrix}^{-1}
  \begin{bmatrix}
    \mat{R}_\mathrm{*K} & \mat{R}_\mathrm{*N}
  \end{bmatrix}\T.
  \label{eqn:postmeancov}
\end{equation}
Because the prior covariance decomposes as $\mat{R}_{**} = \langle \vec{m}\vec{m}\T \rangle + \mat{C}$, the sum of the posterior mean and posterior uncertainty spectra reproduces the prior model for the balanced signal.
At high wavenumbers, the posterior uncertainty spectrum exceeds that of the posterior mean, indicating that the reconstruction becomes uncertainty-dominated below wavelengths of \qty{36}{\kilo\meter}---the effective resolution limit of the extraction (Fig.~\ref{fig:balanced_spectrum}a). 
The steep roll-off of the mean spectrum beyond this scale reflects the inability to reconstruct the small scales in the presence of the noise.
Unlike a low-pass filter, the balanced extraction determines the resolution limit directly from the data and provides explicit uncertainty estimates on the fields as a function of scale.
This resolution estimate is consistent with the vorticity map, which displays features that are of order \qty{10}{\kilo\meter} in scale, 
remembering that a wavelength~$\ell = \qty{36}{\kilo\meter}$ corresponds to an eddy radius of $\ell / 2\pi = \qty{6}{\kilo\meter}$ (Fig.~\ref{fig:balanced_extraction}d).

It is instructive to view the extraction in Fourier space, ignoring any edge effects of the actual sampling pattern and considering the KaRIn case in isolation. The linear Gaussian assumption implies that each Fourier coefficient is independent, and the Bayesian inversion becomes trivial. First considering a one-dimensional line measurement, we find that the posterior mean of the Fourier coefficient for the balanced signal is
\begin{equation}
  \hat m(k) = \frac{B(k)}{B(k) + N(k)} \hat h_\mathrm{K}(k),
  \label{eqn:fourierext}
\end{equation}
where $\hat h_\mathrm{K}$ is the Fourier coefficient from the KaRIn measurement. In this idealized setting, the extraction can thus be viewed as a low-pass filter that is tailored to the statistics of the balanced signal and the noise. The posterior variance of the estimated Fourier coefficient is $N(k) B(k) / [B(k) + N(k)]$, and the variance of the posterior mean is $B^2(k) / [B(k) + N(k)]$. These two spectra cross where $B(k) = N(k)$, such that uncertainty dominates the posterior at scales at which the noise dominates the original signal, as expected. For two-dimensional measurements that similarly extend infinitely, the expressions are the same but involve the radial spectra $B_\mathrm{r}(\kappa)$ and $N_\mathrm{r}(\kappa)$, which can be calculated from $B(k)$ and $N(k)$ using an inverse Abel transform (see Appendix~\ref{sec:smoothing}). The expressions for the variance of the posterior mean and the posterior uncertainty can be turned back into one-dimensional spectra with a forward Abel transform. Compared to the one-dimensional case, this produces a slight shift toward a more favorable ratio between the variance of the posterior mean and the posterior uncertainty (Fig.~\ref{fig:balanced_spectrum}b). For the parameters inferred from the actual KaRIn measurements, the transition to uncertainty-dominated scales shifts from a wavelength of \qty{40}{\kilo\meter}, where the one-dimensional $B(k) = N(k)$, down to \qty{36}{\kilo\meter}, matching the scale inferred from the SWOT data sampling described above (Fig.~\ref{fig:balanced_spectrum}a). While this Fourier view of the extraction is useful to build intuition, we perform the extraction in physical space in practice to optimally handle the edges of the swaths, fill in the nadir gap, and combine the KaRIn and nadir data.

\section{Evaluation with Synthetic SWOT Data} \label{sec:synth}

\begin{figure}[p]
  \centering
  \includegraphics[width=0.9\textwidth]{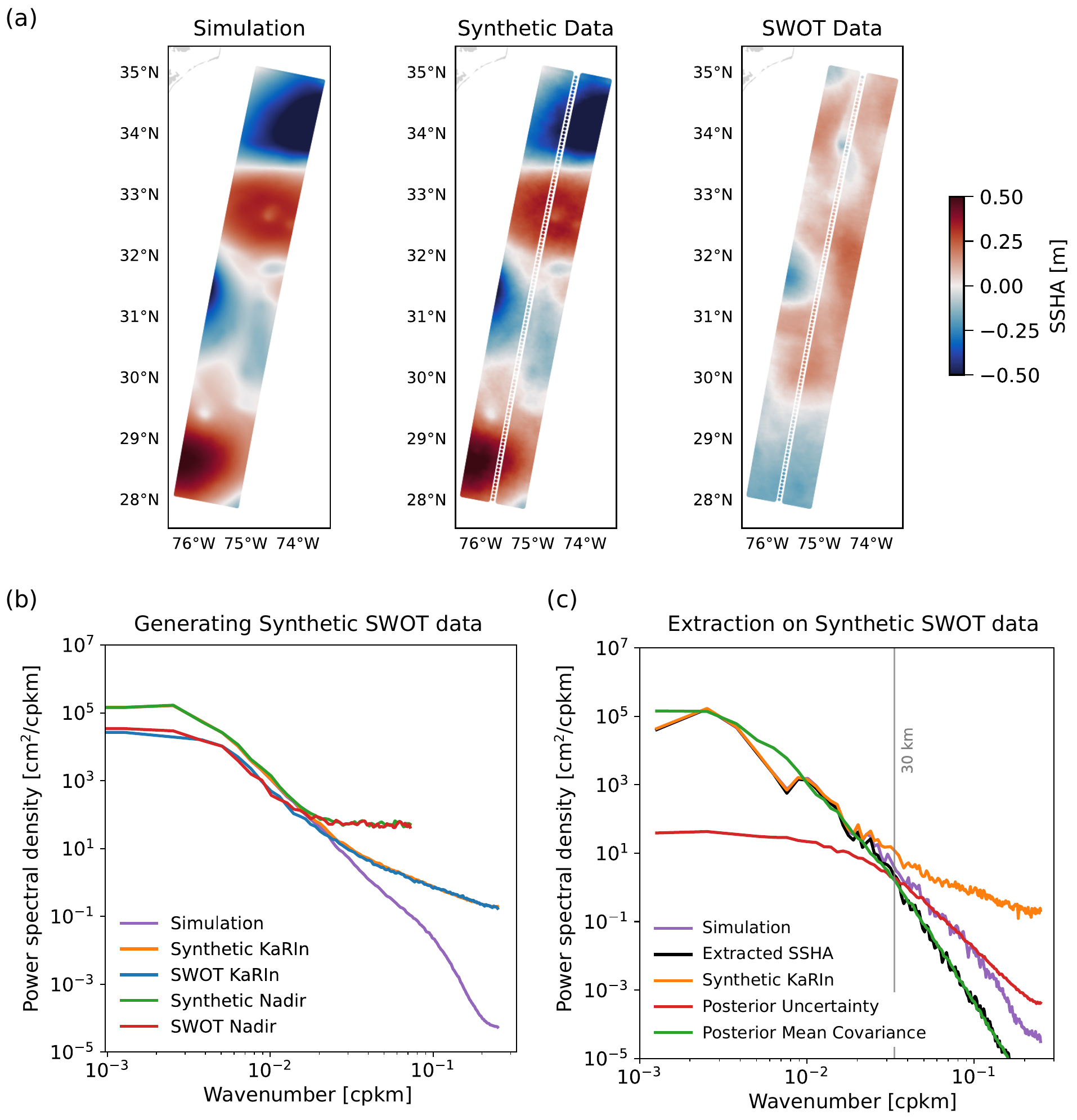}
  \caption{%
    Synthetic SWOT observations and their statistics.
    (a)~Maps of SSHA from the simulation (left), the corresponding synthetic SWOT observations (middle), and the real SWOT measurements (right) along pass~9 on May~10.
    (b)~Time-averaged SSHA variance spectra from the simulation, synthetic data, and real data.
    (c)~SSHA variance spectra from a single snapshot on May~10 from the simulation, the synthetic KaRIn observations, and the SSHA inferred from the synthetic observations by the balanced extraction.
    Also shown are the spectra of the posterior mean and uncertainties, obtained by averaging over samples from the mean covariance~\eqref{eqn:postmeancov} and from the posterior covariance~$\mat{C}$.
    The gray line marks the intersection between these two spectra and indicates the effective resolution of the extraction.
  }
  \label{fig:synthetic_swot_data}
\end{figure}

To evaluate the balanced extraction method, we use synthetic SWOT observations generated from the numerical simulation described in Section~\ref{sec:data}.
The simulation fields serve as a ``ground truth'' against which we can compare the results of adding noise, taking a measurement over the swath geometry, and performing the balanced extraction method. 
All comparisons are based on the simulation SSH field and its spatial derivatives; we make no use of the model velocities.

To generate the synthetic observations, we sample the model fields at the same times of the year as the real SWOT measurements, interpolate them to the exact locations of the KaRIn and nadir data points, and then add realistic noise that matches the spectral characteristics of each instrument.
For the synthetic KaRIn swaths, spatially correlated noise is drawn from the probability distribution characterized by the covariance matrix inferred from the fitted spectral noise model of Section~\ref{sec:spectral_models}. 
Independent realizations of the noise are generated for each SWOT cycle by multiplying vectors with elements drawn from standard normal distributions with the Cholesky factor of this covariance matrix. 
For the synthetic nadir tracks, we add uncorrelated white noise with the standard deviation~$\sigma$ inferred from the spectral fit.
The resulting synthetic SWOT dataset replicates both the sampling geometry of the SWOT measurements and the high-wavenumber spectral characteristics of the SWOT observations, while the original noise-free simulation SSHA fields serve as the benchmark for evaluating the balanced extraction (Fig.~\ref{fig:synthetic_swot_data}a).

The across-track and time-averaged variance spectrum of the simulated SSHA resembles that observed by KaRIn (Fig.~\ref{fig:synthetic_swot_data}b).
The simulation has more variance than the observations at low wavenumbers, but the spectral slope in the submesoscale range, where balanced signals dominate, is comparable.
The simulation, which is free of instrument noise (from here on referred to as ``noiseless''), exhibits a slight spectral hump at around \qty{10}{\kilo\meter} in wavelength, which is due to internal waves \parencite{SkinnerLawrence&Callies_2025}.
We do not attempt to model this internal-wave contribution statistically because it is overwhelmed by noise in both the real and synthetic observations; nevertheless, its presence should be noted when comparing extracted signals with the ground truth.
Overall, the simulation provides a sufficiently realistic test bed for evaluating the balanced extraction.
Furthermore, we note that the simulation supports our balanced extraction model: the balanced region of its spectrum follows a clear power law that extends well below the observation's noise-dominated regime, supporting our assumption that the balanced model can be extrapolated there.

\begin{figure}[p]
  \centering
  \includegraphics[width=\textwidth]{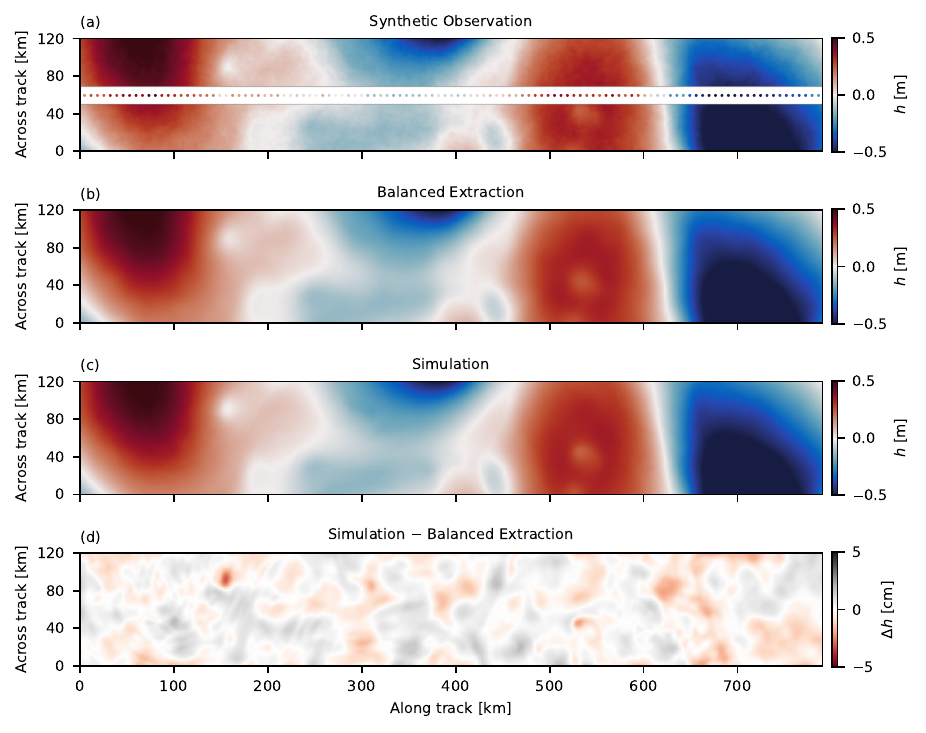}
  \caption{%
    Reconstructing the balanced SSHA field from synthetic SWOT data.
    (a)~Synthetic SWOT observations.
    (b)~Extracted balanced SSHA field, estimated by applying the extraction method to the synthetic observations.
    (c)~The ground truth from the simulation before synthetic noise was added and the SWOT sampling applied.
    (d)~The difference between the balanced extracted fields and the simulation data. Note there is a factor of 10 difference in color scale between panels (a–c) and panel (d).
  }
  \label{fig:ssh_extraction}
\end{figure}

\begin{figure}[p]
  \centering
  \includegraphics[width=0.5\textwidth]{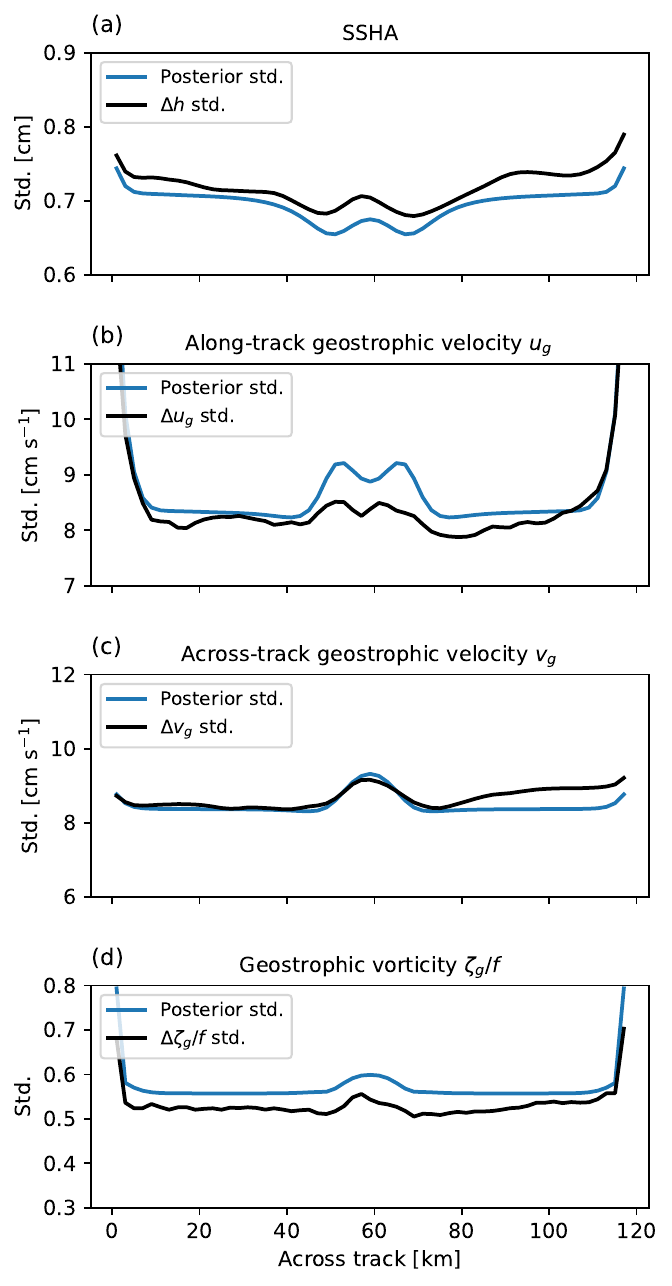}
  \caption{%
    Comparison of posterior uncertainties and errors in the synthetic extraction.
    Shown are the pointwise posterior uncertainties from $\mat{C}$ and the errors in the balanced extraction as a function of across-track distance for SSHA, the two components of the geostrophic velocity, and normalized geostrophic vorticity.
    The errors are averaged over synthetic samples of the simulation corresponding to the rapid-repeat phase of SWOT.
    }
  \label{fig:cross_track_std}
\end{figure}

We first assess the performance of the balanced extraction using synthetic observations of cycle~517 of pass~9 in the same region as discussed above (Fig.~\ref{fig:na_swot_passes}).
We choose this cycle because it is a representative yet geometrically complex test case.
It features vigorous mesoscale eddies and a variety of intricate submesoscale frontal and eddy features (Fig.~\ref{fig:ssh_extraction}c).
The synthetic SWOT data, used as the input field for the extraction, exhibit noticeable small-scale noise and include the central nadir gap, reproducing the characteristic features of SWOT observations (Fig.~\ref{fig:ssh_extraction}a).
Despite the small-scale noise and nadir gap, the balanced extraction produces an accurate reconstruction of the original SSHA field (Fig.~\ref{fig:ssh_extraction}b).
The comparison between the reconstructed field and the original input shows differences with a root-mean-square (RMS) of \qty{0.7}{\centi\meter} (Fig.~\ref{fig:ssh_extraction}d and Fig.~\ref{fig:cross_track_std}a).
Two submesoscale cyclones at $x = \qtylist{160;520}{\kilo\meter}$ and the edge of a mesoscale cyclone around $x = \qty{390}{\kilo\meter}$ are slightly suppressed in amplitude and smoothed in their structure, but they are clearly captured by the reconstruction.
The extraction also seamlessly interpolates across the nadir gap; it reconnects the eddy structures and fronts that were fragmented by the KaRIn and nadir sampling in the observations.
The balanced extraction accurately recovers the underlying ground truth signal in the gap without introducing any spurious features.

As with the real SWOT data, the uncertainty is dominated by the small scales, where noise obscures the signal (Fig.~\ref{fig:synthetic_swot_data}c).
The extracted balanced signal has a spectrum that closely follows the spectrum of the ground truth at large scales but starts rolling off more steeply at around \qty{30}{\kilo\meter}.
The spectrum of the posterior uncertainty exceeds that of the posterior mean at wavelengths below an effective resolution of $\ell = \qty{30}{\kilo\meter}$.
This corresponds to a resolvable feature scale of some $\ell / 2\pi = \qty{5}{\kilo\meter}$---i.e., a Gaussian eddy diameter of $\ell/\pi = \qty{10}{\kilo\meter}$.
This transition occurs at slightly smaller scales than in the real case because the balanced signal is somewhat stronger in the simulation than in the real ocean (Fig.~\ref{fig:synthetic_swot_data}b).

To evaluate the accuracy of the uncertainty estimate of the balanced extraction, in the face of the imperfect statistical description of the balanced signal, we compare the RMS differences between the extracted balanced fields and the (noiseless) simulation fields with the posterior standard deviation of the balanced extraction (Fig.~\ref{fig:cross_track_std}).
The RMS differences between the extracted balanced fields and the simulation fields are averaged both along the track and over all samples in the rapid repeat phase.
The posterior standard deviation of the extraction, computed as the square root of the diagonal elements of~$\mat{C}$, as above, is also averaged in the along-track direction, although there is little along-track variation.

\begin{figure}[p]
  \centering
  \includegraphics[width=\textwidth]{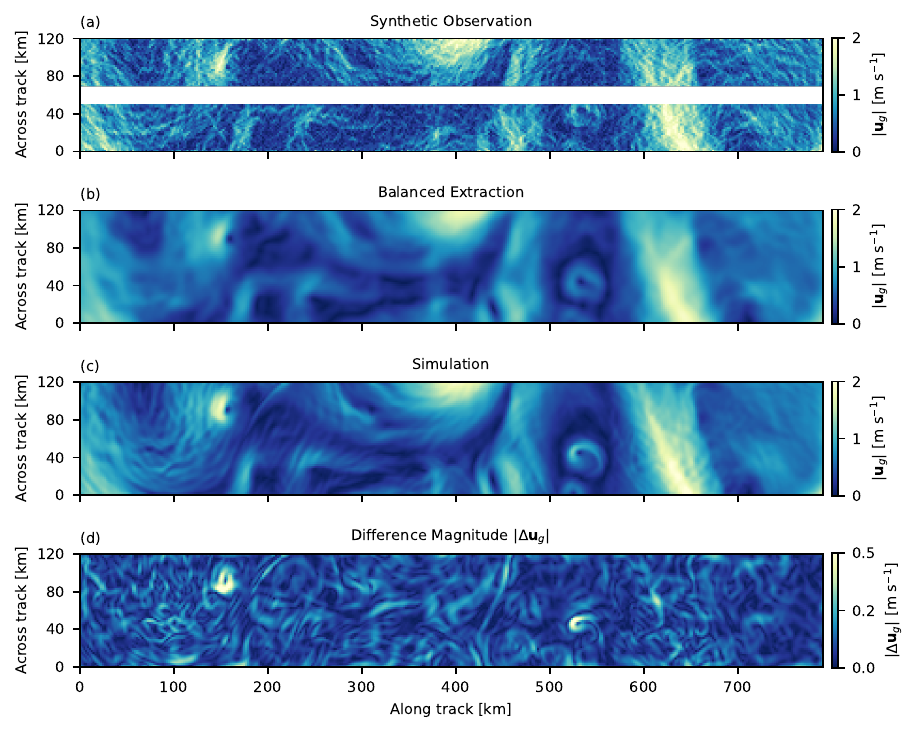}
  \caption{%
    Evaluation of the geostrophic speed inferred from synthetic observations.
    (a)~Geostrophic speed calculated directly from the synthetic KaRIn observations.
    (b)~Geostrophic speed from the balanced SSHA field extracted from the synthetic observations.
    (c)~Ground truth geostrophic speed from the simulation.
    (d)~Magnitude of the difference in the geostrophic velocities from the balanced extraction and the ground truth.
    Note there is a factor of $4$ difference in color scale between panels (a--c) and (d). 
  }
  \label{fig:grad_extraction}
\end{figure}

For the SSHA fields, the posterior standard deviation estimate in the across-track direction ranges from \qty{0.70}{\centi\meter} in the centers of the KaRIn swaths to \qty{0.68}{\centi\meter} within the nadir gap (Fig.~\ref{fig:cross_track_std}a).
This posterior uncertainty aligns very closely with the RMS differences in both magnitude and across-swath structure (within $\sim$5\%).
The same comparison for the geostrophic velocity components and the geostrophic vorticity, whose uncertainty depends on error covariances in the height field, further supports the use of the posterior uncertainty as a reasonable estimate of errors in the extracted signal (Fig.~\ref{fig:cross_track_std}b--d).
The along-track velocity~$u_g$ posterior standard deviation ranges from \qty{8.3}{\centi\meter\per\second} in the centers of the KaRIn swaths to \qty{12.5}{\centi\meter\per\second} at the edges of the swaths, which is very close to, but a slight overestimate of, the RMS differences that range from \qtyrange{8.1}{12.4}{\centi\meter\per\second} (Fig.~\ref{fig:cross_track_std}b). 
The estimates of uncertainty for the across-track component~$v_g$ match almost exactly, with a posterior standard deviation ranging from \qty{8.3}{\centi\meter\per\second} in the centers of the KaRIn swaths to \qty{9.3}{\centi\meter\per\second} in the nadir gap, compared with RMS differences ranging from \qtyrange{8.3}{9.2}{\centi\meter\per\second} and displaying a similar structure with a noticeable increase in the nadir gap (Fig.~\ref{fig:cross_track_std}c).
For context, these velocity uncertainties are small compared to the balanced signal, which reaches magnitudes of \qtyrange{1}{2}{\meter\per\second} (Fig.~\ref{fig:grad_extraction}b).
For the geostrophic vorticity, the posterior standard deviation ranges from $0.55f$ in the centers of the KaRIn swaths to $0.60f$ in the nadir gap, compared with RMS differences ranging from $0.52f$ in the centers of the KaRIn swaths to $0.54f$ in the nadir gap (Fig.~\ref{fig:cross_track_std}d).
The posterior uncertainty is slightly larger than the RMS difference, which is expected because the balanced model extrapolates the power law to high wavenumbers ad~infinitum, whereas the simulation has finite resolution and deviates substantially from this power law in its dissipation range at wavelengths \qty{<10}{\kilo\meter} (Fig.~\ref{fig:synthetic_swot_data}b,c).
As a result, the simulation has less geostrophic enstrophy than predicted by the balanced model, and the RMS differences are somewhat smaller than predicted.
The geostrophic vorticity statistics are discussed further below.
Overall, the posterior uncertainties align well with the true reconstruction errors, demonstrating that the posterior covariance provides a good estimate of the true reconstruction uncertainty and captures the spatial variations in uncertainty across the swaths and nadir gap.

The removal of the noise by the balanced extraction is particularly important for derived dynamical quantities such as geostrophic velocity and vorticity.
The geostrophic speed computed from the raw synthetic observations displays high-amplitude striated patterns that appear to be associated with the larger-scale structures, obscuring the finer-scale fronts and eddy peripheries (Fig.~\ref{fig:grad_extraction}a).
This association is completely spurious, given that the synthetic noise is uncorrelated to the signal.
The apparent association between the noise and signal is a result of the geostrophic speed being a nonlinear diagnostic.
In some regions, there are weaker signals with a similar pattern in the simulation (Fig.~\ref{fig:grad_extraction}c), which appear to be due to internal waves (see the spectral hump in Fig.~\ref{fig:synthetic_swot_data}b).
Extreme caution should be exercised in interpreting such patterns in the real SWOT data, and the spurious striations due to the noise should not be confused with the much weaker real signal predicted by the simulation.
The balanced extraction effectively removes the dominant noise artifacts, yielding a smooth geostrophic speed field and seamlessly filling the nadir gap (Fig.~\ref{fig:grad_extraction}b).
The extracted field is smoother than the ground truth, as expected from the spectral steepening of the expected signal at the uncertainty-dominated scales (Fig.~\ref{fig:synthetic_swot_data}c).

\begin{figure}[p]
  \centering
  \includegraphics[width=0.89\textwidth]{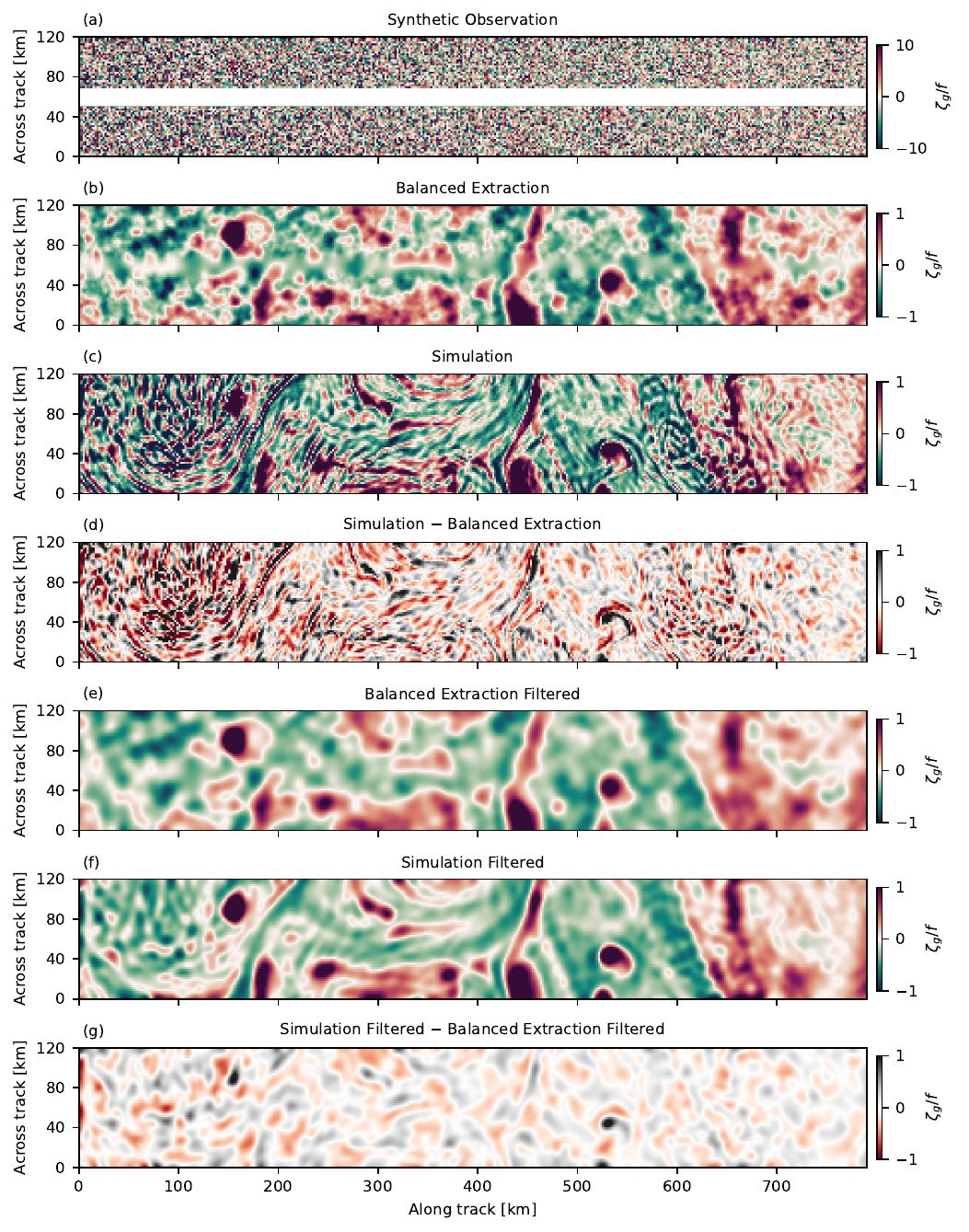}
  \caption{%
    Evaluation of the geostrophic vorticity inferred from synthetic observations.
    (a)~Geostrophic vorticity calculated directly from the synthetic KaRIn observations.
    (b)~Geostrophic vorticity from the balanced SSHA field extracted from the synthetic observations.
    (c)~Ground truth geostrophic vorticity from the simulation.
    (d)~Difference in the geostrophic vorticity from the balanced extraction and the ground truth.
    (e)~Geostrophic vorticity from the balanced extraction and filtered to \qty{4}{\kilo\meter}.
    (f)~Ground truth geostrophic vorticity from the simulation, filtered to \qty{4}{\kilo\meter}.
    (g)~Difference in the geostrophic vorticity from the filtered balanced extraction and the filtered ground truth.
  }
  \label{fig:vort_extraction}
\end{figure}

The small-scale noise in the synthetic observations is further amplified in the geostrophic vorticity field, rendering the noise far greater than the signal in the raw fields (Fig.~\ref{fig:vort_extraction}a).
In contrast, the simulation's geostrophic vorticity field is dominated by energetic submesoscale cyclones with magnitudes exceeding the local planetary vorticity~$f$, along with elongated filaments and fronts (Fig.~\ref{fig:vort_extraction}c).
Some contribution from internal waves is also evident at the smallest resolved scales.
The balanced extraction successfully recovers a smoothed version of this ground truth simulation field (Fig.~\ref{fig:vort_extraction}b): the intense cyclones, major fronts, and filaments are all retained, although slightly blurred, especially in the nadir gap. 
Note that the real SWOT vorticity fields obtained from the balanced extraction also follow this same overall structure (Fig.~\ref{fig:balanced_extraction}d).
Because vorticity depends on small-scale gradients, the reconstruction carries a larger relative uncertainty, with a posterior standard deviation of approximately~$0.6f$ (Fig.~\ref{fig:cross_track_std}d), consistent with the RMS difference between the extracted and ground truth fields (Fig.~\ref{fig:vort_extraction}d).

According to the statistical model, the balanced height field is only once differentiable: its spectral slope $s_\mathrm{b} \le 5$ causes the variance of its second derivative to diverge \parencite[e.g.,][]{adler_geometry_2010}. As a result, the geostrophic vorticity calculated by finite differences is dominated by the smallest resolved scales.
The power in the balanced signal extracted from the noisy data is attenuated at small scales (Fig.~\ref{fig:synthetic_swot_data}c), and the geostrophic vorticity field is dominated by features near the scale at which the noise starts having more power than the signal (Fig.~\ref{fig:vort_extraction}b).
For the ground truth simulation data, the model resolution sets the scale dominating the geostrophic vorticity field (Figs.~\ref{fig:synthetic_swot_data}c,~\ref{fig:vort_extraction}c).
To allow for a more meaningful assessment of the extracted geostrophic vorticity field, we therefore compare the extraction of a filtered balanced signal to that of a filtered ground truth.
We apply a Gaussian filter with a smoothing scale~$\rho$ to the target signal, such that the target fields are infinitely differentiable.
The covariance matrices involving the target now read
\begin{align}
  (R_\mathrm{**})_{ij} &= \mathcal{C}[\mathcal{A}[\mathcal{A}^{-1}[B(k)](\kappa) \exp(-\rho^2 \kappa^2)](k)](r), \label{eqn:smoothing-1} \\
  (R_\mathrm{*N})_{ij} &= \mathcal{C}[\mathcal{A}[\mathcal{A}^{-1}[B(k)](\kappa) \exp(-\tfrac{1}{2} \rho^2 \kappa^2)](k)](r), \label{eqn:smoothing-2} \\
  (R_\mathrm{*K})_{ij} &= \mathcal{C}[\mathcal{A}[\mathcal{A}^{-1}[B(k)](\kappa) \exp(-\tfrac{1}{2}\rho^2 \kappa^2)](k)](r), \label{eqn:smoothing-3}
\end{align} 
where $\mathcal{A}$ denotes an Abel transform and $\mathcal{A}^{-1}$ its inverse, which are needed because the smoothing is applied in two dimensions and the spectra are one-dimensional (see Appendix~\ref{sec:smoothing}).
To ensure a like-for-like comparison, we process the synthetic SWOT observations and the noiseless simulation fields using the same balanced-extraction framework.
For the noiseless version, we sample the simulation at the full set of target locations, including in the nadir gap, and apply no noise in the data covariance.
We set $\rho/2\pi = \qty{4}{\kilo\meter}$, roughly corresponding to the effective resolution of the synthetic KaRIn data inferred above (Fig.~\ref{fig:synthetic_swot_data}c).
As expected, the correspondence between the extracted signal and the ground truth is much improved by the smoothing (Fig.~\ref{fig:vort_extraction}e,f).
Many of the submesoscale eddies, fronts, and filaments at this smoothing scale are captured by the extraction.
The RMS difference is reduced to $0.18f$ (Fig.~\ref{fig:vort_extraction}g).

A particular application of our balanced extraction technique is to compute normalized histograms of the pointwise geostrophic vorticity.
This statistical characterization of the flow provides a fundamental diagnostic of balanced turbulence.
It quantifies the intensity of mesoscale and submesoscale eddies, reveals asymmetries between cyclonic and anticyclonic motions, and characterizes the structure of eddy--eddy interactions, intermittency, and coherent vortices that mediate energy transfer across scales \parencite[e.g.,][]{McWilliams_1984, McWilliams_1985, Chelton_etal_2011, Shcherbina_etal_2013}.
Accurately recovering these statistics from SWOT observations is therefore very useful for testing competing theories of turbulence, constraining the energy transfer at meso- and submesoscales, and evaluating the realism of ocean circulation models.
Vorticity histograms are expected to be skewed and have wide tails, such that it is particularly important that they are evaluated carefully in order to ensure that data processing does not distort them in unrecognized ways \parencite[e.g.,][]{Baker_etal_1987,Sardeshmukh_etal_2015,Gaultier_etal_2016,Martin_etal_2017}.
To this end, calculating the vorticity histograms from the balanced extraction in the simulation, where we can also compute the ground-truth histograms, is an important check on the reliability of such statistics in real data.

\begin{figure}[t]
  \centering
  \includegraphics[width=\textwidth]{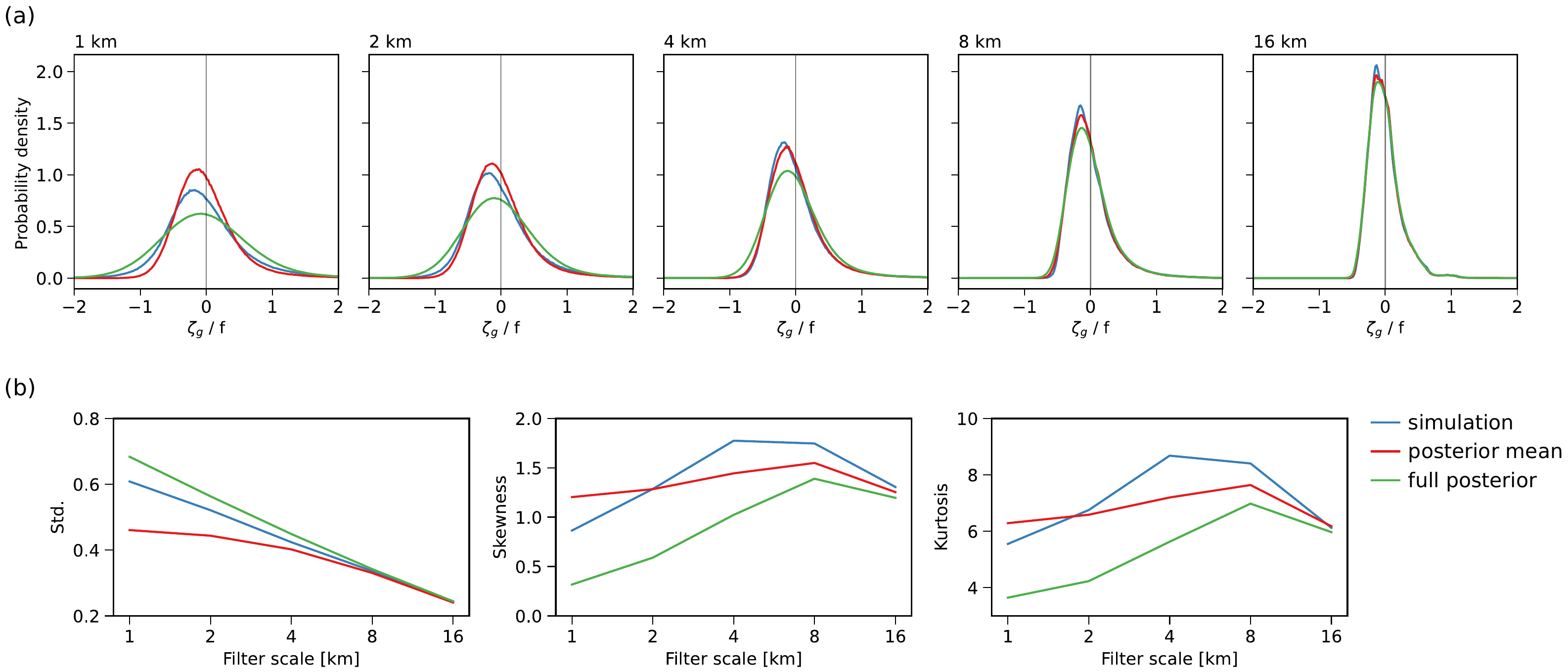}
  \caption{%
    Statistics of the geostrophic vorticity $\zeta_g$ from the simulation and balanced extraction at different smoothing scales.
     (a)~Normalized histograms of geostrophic vorticity~$\zeta_g / f$, averaged over the rapid-repeat phase.
     Shown are histograms for the ground-truth simulation fields (blue), the balanced fields extracted from the synthetic SWOT observations (posterior mean, red), and samples from the full posterior distribution (green).
     Curves are shown for progressively increasing spatial smoothing scales from \qtyrange{1}{16}{\kilo\meter}.
     (b)~The standard deviation, skewness, and kurtosis of the vorticity distributions as a function of smoothing scale. 
  }
  \label{fig:pdf_cyclon}
\end{figure}

We apply the extraction with varying smoothing scales $\rho/2\pi = \qtylist{1;2;4;8;16}{\kilo\meter}$, and we compute normalized histograms of the geostrophic vorticity~$\zeta_g/f$ from the smoothed ground truth, from the extracted fields (the posterior mean), and from 20~samples drawn from the full posterior distribution, with statistics accumulated over all samples of the rapid-repeat phase.
As expected, the histograms of the balanced extraction align closely with those of the ground truth for the most smoothed fields, and they progressively diverge as the smoothing scale is reduced (Fig.~\ref{fig:pdf_cyclon}).
As the smoothing scale drops below the effective resolution of the balanced extraction, the extracted signal has a substantially reduced variance because it does not capture the small scales that are dominated by noise in the synthetic observations.
This additional variance is captured by the posterior uncertainty; in fact, the posterior histogram has slightly too much variance because the balanced model has somewhat more variance at the smallest scales resolved by the simulation, as discussed above (Fig.~\ref{fig:synthetic_swot_data}b,c).

Despite the assumption of Gaussian statistics in the extraction method, the extracted signal captures some of the non-Gaussian character of the underlying geostrophic vorticity field where the data constraints are strong (Fig.~\ref{fig:pdf_cyclon}).
The simulation has a positive skewness of \num{1.8} when filtered to \qtylist{4}{\kilo\meter}, and the extracted signal (the posterior mean) has a skewness of \num{1.4}.
The full posterior histogram is less skewed, with a value of \num{1.0}, because the part of the signal that is not constrained by the data is filled in with a Gaussian process, pushing the posterior histogram toward a normal distribution.
The same effect is exhibited by the kurtosis: the simulation has wide tails, which is captured to a degree by the balanced extraction.
At \qty{4}{\kilo\meter}, the kurtosis is \num{8.6} for the simulation, \num{7.2} for the extracted signal (posterior mean), and \qty{5.5} for the posterior histogram. 
As the smoothing scale is reduced, the simulation becomes more Gaussian, presumably due to the emergence of an internal-wave signal at the smallest resolved scales (Fig.~\ref{fig:vort_extraction}c).
The extracted signal does not capture this behavior because these scales are severely under-resolved, and the statistics are dominated by the more non-Gaussian larger scales.
As a result, both skewness and kurtosis are overestimated (Fig.~\ref{fig:pdf_cyclon}).
The posterior histogram, which fills in the missing information with a Gaussian signal, instead provides a lower bound on the skewness and kurtosis.

\section{Discussion}
\label{sec:discussion}

To study balanced meso- and submesoscale turbulence using SWOT data, their SSHA signature must be isolated from other components of the signal.
The method presented here separates the observed signal into two components, one due to balanced flows and one due to small-scale noise, under the assumption that the balanced SSHA spectrum has a steeper spectral roll-off than the other contributors to the spectrum.
We developed a Bayesian inversion that models these two components as Gaussian processes with distinct statistics inferred from wavenumber spectra. 
This spectral model provides the prior, which the inversion combines with the observations to obtain a posterior estimate of the balanced signal.

This inversion provides reliable best estimates as well as uncertainties for the balanced part of the signal, as demonstrated by analyzing synthetic observations of simulation output.
In the Northwest Atlantic region considered here, the inversion produces an estimate of the balanced signal with an effective resolution of $\ell = \qty{36}{\kilo\meter}$, corresponding to a feature scale of $\ell / 2\pi = \qty{6}{\kilo\meter}$, enabling it to capture submesoscale eddies, fronts, and filaments.
This resolution scale depends on the relative amplitude of the balanced signal and noise and will therefore vary across the world ocean.

The present extraction method assumes that the signal has no substantial contribution from internal tides, which limits its applicability to the $\sim$30\% of the global ocean where balanced flows dominate, such as western boundary current regions and the Southern Ocean.
\textcite{Zhang_2025} estimated that low-mode internal tides contribute substantially to the SWOT signal in $\sim$70\% of the regions sampled during the rapid-repeat phase.
The present method can be extended to allow for such signals by adding more terms to the covariance function.
These additional terms would aim to characterize the spatial structure of the additional signals, and they would allow the inversion to distinguish tidal signals from the rest of the signal to the degree that their spatial structure is distinct.
Given the narrowband nature of low-mode internal tides \parencite[e.g.,][]{ray_m2_2016}, this approach should be able to separate them from the broadband balanced signal.
Taking advantage of the tides' narrowband nature in the frequency domain and applying the extraction simultaneously to multiple cycles with aliased tidal signals could further improve estimates.
Other approaches such as harmonic fits \parencite[e.g.,][]{zaron_baroclinic_2019}, neural nets trained on simulations \parencite[e.g.,][]{wang_deep_2022,xiao_reconstruction_2023,gao_deep_2024}, or model-based predictions \parencite{yadidya_advancing_2025} could also be used to remove the internal tide before further analysis of the balanced signal.

Our method further assumes that the flow statistics are stationary over the averaging period. 
This is a reasonable assumption for the balanced signal during the short timeframe of the rapid-repeat phase used here.
However, during the science phase of the SWOT mission, there is good evidence that both the small-scale signal and the balanced component will have seasonally variable statistics \parencite{callies_2015, buckingham_etal_2016, rocha_2016, lawrence_2022, Zhang_2025}.
This can be addressed by estimating spectra and covariance parameters within seasonal subsets of the data, such that the flow statistics remain effectively stationary within each analysis period.
The small-scale noise component also exhibits some spatial and temporal non-stationarity.
It increases in amplitude toward the edges of the swaths \parencite{kachelein_sub-100_2025} and follows day-to-day variations in the significant wave height \parencite{Zhang_2025}.
These variations of the noise in space and time can be incorporated into the noise covariance function if sufficiently well characterized.

Another limitation of the present approach is that it models the balanced signal as a Gaussian process, ignoring the known non-Gaussian statistics of balanced meso- and submesoscale turbulence.
While the present method does recover some of these non-Gaussian statistics where the data constraint is strong, employing a non-Gaussian prior might help better preserve sharp fronts and filaments in the estimate.
Approaches that learn such non-Gaussian statistics from simulation output may produce sharper submesoscale features \parencite[e.g.,][]{treboutte_karin_2023,febvre_training_2024,martin_generative_2025}.
However, the degree to which the simulation output is consistent with the SWOT data is not currently understood, and the potential for model error biasing the estimates should be kept in mind in applications.

\section{Conclusions}

The proposed extraction method transparently produces robust estimates of the balanced signal with explicit uncertainties.
This enables the study of the dynamics of meso- and submesoscale turbulence, provided balanced flows are energetic and the internal tides are sufficiently weak.
In the Gulf Stream region studied here, the balanced signal can be extracted with confidence down to a feature scale of about \qty{6}{\kilo\meter}, as determined by the posterior uncertainty.
The method also provides a framework for evaluating the realism of simulation output, to understand how accurately simulations capture the physics of meso- and submesoscale turbulence.
The method is extensible and can be adapted to include more refined dynamical assumptions, such as additional signal components, non-stationary statistics, and region-specific priors.

%
%

\section*{Open Research}

For the SWOT data, we use the Level-2 KaRIn Basic Low Rate SSH v2.0 (\url{https://doi.org/10.5067/SWOT-SSH-2.0}) and SWOT Level-2 Nadir Altimeter Geophysical Data Record (GDR) with Waveforms (\url{https://doi.org/10.5067/SWOT-NALT-GDR-2.0}), which are available from NASA PO.DAAC and Earthdata \parencite{KaRIn, Nadir}.
For the simulations outputs, we use the model described by \textcite{Sinha_2023,SkinnerLawrence&Callies_2025}.
The analysis code for spectral fitting and Gaussian-process reconstruction, along with figure scripts, is hosted on GitHub (\url{https://github.com/jwskinner/SWOT-balanced-extraction}).

\section*{Acknowledgments}

\sloppy

This work was supported by NASA grants 80NSSC23K0345, 80NSSC20K1140, and 80NSSC24K1652, by a Simons Foundation Pivot Fellowship, a Simons Foundation Pivot Research Award, and by NSF grant PHY-2309135 to the Kavli Institute for Theoretical Physics (KITP).
This work used high-performance computing at the San Diego Supercomputer Center \parencite{SDSC2022} through allocation PHY-230189 from The Advanced Cyberinfrastructure Coordination Ecosystem: Services \& Support (ACCESS) program, which is supported by National Science Foundation grants \#2138259, \#2138286, \#2138307, \#2137603, and \#2138296.
The authors thank Patrice Klein, Andrew Thompson, Dimitris Menemenlis, Hector Torres, and Jinbo Wang for helpful discussions, as well as Gregg Jacobs and two anonymous reviewers for helpful comments.

\fussy

\appendix

\section{Two-dimensional smoothing}
\label{sec:smoothing}

In this section, we describe the formulae for the covariances of the smoothed fields used in Section~\ref{sec:synth}. 
We estimate the one-dimensional spectra from the observations. 
These are related via a cosine transform to the covariance function used in the extraction.
However, the smoothing is applied in two-dimensional spectral space, whereas the observations provide only one-dimensional spectra. Consequently, we must first transform the observed one-dimensional spectra into a two-dimensional spectral representation, apply the smoothing, and then compute the corresponding covariances.

Assuming isotropy, a one-dimensional spectrum~$P(k)$ can be converted to the two-dimensional spectrum~$P_2(k, l)$ and the radial spectrum $P_\mathrm{r}(\kappa)$. We normalize the spectra such that
\begin{equation}
  \int_0^\infty P(k) \, \d k = \int_{-\infty}^\infty \int_{-\infty}^\infty P_2(k, l) \, \d k \, \d l = \int_0^\infty P_\mathrm{r}(\kappa) \, \d \kappa.
\end{equation}
Because $\d k \, \d l = \kappa \, \d \kappa \, \d \theta$, the two-dimensional and radial spectra are related by
\begin{equation}
  2\pi \kappa P_2(k, l) = P_\mathrm{r}(\kappa).
\end{equation}
The one-dimensional spectrum is (twice) the integral of the two-dimensional spectrum over one dimension, such that the one-dimensional spectrum can be related to the radial spectrum as follows:
\begin{equation}
  P(k) = 2 \int_{-\infty}^\infty P_2(k, l) \, \d l = \frac{2}{\pi} \int_0^\infty \frac{P_\mathrm{r}(\kappa)}{\kappa} \, \d l = \frac{2}{\pi} \int_k^\infty \frac{P_\mathrm{r}(\kappa)}{\sqrt{\kappa^2 - k^2}} \, \d \kappa \equiv \mathcal{A}[P_\mathrm{r}(\kappa)](k).
\end{equation}
This is an Abel transform, and the inverse Abel transform gives
\begin{equation}
  P_\mathrm{r}(\kappa) = -\kappa \int_\kappa^\infty \frac{P'(k)}{\sqrt{k^2 - \kappa^2}} \, \d k \equiv \mathcal{A}^{-1}[P(k)](\kappa),
\end{equation}
where the prime denotes a derivative. This allows us to go back and forth between one-dimensional spectra and radial spectra.

To represent the spatial smoothing, we apply an inverse Abel transform to $B(k)$ from~\eqref{eq:balanced_model} to calculate the two-dimensional spectrum of the balanced signal, which we then multiply by the Gaussian smoothing $\exp(-\frac{1}{2} \rho^2 \kappa^2)$ before applying the Abel transform to go back to a one-dimensional spectrum, resulting in~\eqref{eqn:smoothing-1}--\eqref{eqn:smoothing-3}. 
This recipe can be used for the covariance between any fields, with any smoothing applied to them.

While some Abel transforms can be done analytically, we implement the forward transform~$\mathcal{A}$ by discretely sampling the spectrum at a uniform wavenumber grid from $0$ to $n/2L$ with spacing~$L^{-1}$, approximating $P_\mathrm{r}(\kappa)$ as linear between sampling points and then evaluating the integral analytically for each sampling interval. For the inverse transform~$\mathcal{A}^{-1}$, we first calculate $P'(k)$ using second-order accurate finite differences on the same wavenumber grid before following the same approach as for the forward transform. The cosine transform~$\mathcal{C}$ to obtain the covariance function is performed using the type-I discrete cosine transform \parencite[DCT-I; e.g.,][]{Strang_1999} on the same wavenumber grid. In all calculations, we use $L = \qty{5000}{\kilo\meter}$ and $n = \num{100000}$.

\end{document}